\documentclass[journal]{IEEEtran}
\usepackage{acronym}
\IEEEoverridecommandlockouts
\usepackage{cite}
\usepackage{amsmath,amssymb,amsfonts}
\usepackage{algorithmic}
\usepackage{comment}
\usepackage{graphicx}
\usepackage{authblk}
\usepackage[caption=false,font=footnotesize]{subfig}
\usepackage{multirow}
\usepackage{textcomp}
\usepackage{flushend}
\usepackage{amssymb}
\usepackage{xcolor}
\usepackage{pgfplots}
\usepackage{microtype}
\usepackage{algorithm}      
\usepackage{fancyhdr}
\usepackage{enumitem}
\usepackage{tikz}
\usepackage{amsthm}

\newcommand{\N}[0]{\ensuremath{\mathcal{N}}}

\newcommand{\Mod}[1]{\, \mathrm{mod} \, #1}

\DeclareMathOperator{\atan2}{atan2}
\DeclareMathOperator{\argmin}{arg\,min\,}

\DeclareMathOperator{\diag}{diag}
\DeclareMathOperator{\blkdiag}{blkdiag}

\newcommand{\appropto}[0]{%
    \ensuremath{%
        \mathrel{%
            \vcenter{%
                \offinterlineskip\halign{%
                    \hfil$##$\cr%
                    \propto\cr\noalign{\kern2pt}%
                    \sim\cr\noalign{\kern-2pt}%
                }%
            }%
        }%
    }%
}

\renewcommand{\vec}[1]{\ensuremath{{\mathbf{#1}}}}
\newcommand{\vx}[0]{\vec{x}}
\newcommand{\vs}[0]{\vec{s}}

\newcommand{\vI}[0]{\vec{I}}
\newcommand{\vh}[0]{\vec{h}}
\newcommand{\vJ}[0]{\vec{J}}

\newcommand{\va}[0]{\vec{a}}

\newcommand{\ve}[0]{\vec{e}}
\newcommand{\vt}[0]{\vec{t}}

 \newcommand{\vm}[0]{\vec{m}}
  \newcommand{\vw}[0]{\vec{w}}
\newcommand{\vp}[0]{\vec{p}}

\newcommand{\vW}[0]{\vec{W}}
\newcommand{\vz}[0]{\vec{z}}
\newcommand{\vA}[0]{\vec{A}}
\newcommand{\vB}[0]{\vec{B}}
\newcommand{\vC}[0]{\vec{C}}

\newcommand{\vE}[0]{\vec{E}}

\newcommand{\vP}[0]{\vec{P}}
\newcommand{\vH}[0]{\vec{H}}
\newcommand{\vR}[0]{\vec{R}}

\newcommand{\vecsymbol}[1]{\ensuremath{\boldsymbol{#1}}}

\newcommand{\vphi}[0]{\vecsymbol{\phi}}
\newcommand{\vpsi}[0]{\vecsymbol{\psi}}
\newcommand{\vtheta}[0]{\vecsymbol{\theta}}

\newcommand{\vOmega}[0]{\vecsymbol{\Omega}}

\newcommand{\cI}[0]{\mathcal{I}}
\newcommand{\cZ}[0]{\mathcal{Z}}
\newcommand{\cP}[0]{\mathcal{P}}

\newcommand{\cM}[0]{\mathcal{M}}

\acrodef{SLAM}[SLAM]{simultaneous localization and mapping}
\acrodef{IP}[IP]{incidence point}

\acrodef{TX}[TX]{transmitter}
\acrodef{RX}[RX]{receiver}
\acrodef{UE}[UE]{user equipment}
\acrodef{BS}[BS]{base station}
\acrodef{TOA}[ToA]{time-of-arrival}
\acrodef{AOD}[AoD]{angle-of-departure}
\acrodef{AOA}[AoA]{angle-of-arrival}
\acrodef{LOS}[LoS]{line-of-sight}
\acrodef{NLOS}[NLoS]{non-line-of-sight}
\acrodef{SVD}[SVD]{singular value decomposition}
\acrodef{LS}[LS]{least squares}
\acrodef{CPD}[CPD]{change point detection}
\acrodef{RMSE}[RMSE]{root mean squared error}
\acrodef{RANSAC}[RANSAC]{random sample consensus}
\acrodef{TPR}[TPR]{true positive rate}
\acrodef{TNR}[TNR]{true negative rate}
\acrodef{MLE}[MLE]{maximum likelihood estimation}
\acrodef{MAP}[MAP]{maximum a posteriori}
\acrodef{CDF}[CDF]{cumulative distribution function}
\acrodef{ISAC}[ISAC]{integrated sensing and communication}
\acrodef{BP}[BP]{belief propagation}
\acrodef{mmwave}[mmWave]{millimeter wave}
\acrodef{IP}[IP]{incidence point}
\acrodef{SP}[SP]{scattering point}
\acrodef{DA}[DA]{data association}
\acrodef{FIM}[FIM]{Fisher information matrix}
\acrodef{EFIM}[EFIM]{equivalent Fisher information matrix}
\acrodef{CRB}[CRB]{Cram$\acute{\text{e}}$r-Rao bound}
\acrodef{MSE}{mean square error}

\acrodef{PEB}[PEB]{position error bound}
\acrodef{LEB}[LEB]{landmark error bound}
\acrodef{HEB}[HEB]{heading error bound}
\acrodef{BEB}[BEB]{clock bias error bound}
\acrodef{OEB}[OEB]{orientation error bound}

\acrodef{OFDM}[OFDM]{orthogonal frequency division multiplexing}
\acrodef{UPA}[UPA]{uniform planar array}

\acrodef{EKF}[EKF]{extended Kalman filter}
\acrodef{PF}[PF]{particle filter}
\acrodef{PRS}[PRS]{positioning reference signal}

\theoremstyle{remark}

\newenvironment{psmallmatrix}
{\left[\begin{smallmatrix}}
{\end{smallmatrix}\right]}

\ifCLASSINFOpdf
\else
\fi

\newcommand{\redcross}{\tikz[baseline]{\draw[red,solid,line width = 1pt] 
(0.0mm,0.0mm) -- (1.6mm,1.6mm) (0.0mm,1.6mm) -- (1.6mm,0.0mm)}}
\newcommand{\redtriangle}{\tikz[baseline]{\draw[red,solid,line width = 1pt] (0.0mm,0.0mm) -- (2.0mm,0.0mm) --  (1.0mm,1.6mm) -- (0.0mm,0.0mm)}}
\newcommand{\redupsidedowntriangle}{\tikz[baseline]{\draw[red,solid,line width = 1pt] (0.0mm,1.6mm) -- (2.0mm,1.6mm) --  (1.0mm,0.0mm) -- (0.0mm,1.6mm)}}

\allowdisplaybreaks

\begin{document}
\bstctlcite{IEEEexample:BSTcontrol}   
%
\title{Exploiting Double-Bounce Paths in Snapshot\\Radio SLAM: Bounds, Algorithms and Experiments}
%
%
%


\author{Xi~Zhang,~\IEEEmembership{Student Member,~IEEE,}
Yu~Ge,~\IEEEmembership{Member,~IEEE,} 
    Ossi~Kaltiokallio,~\IEEEmembership{Member,~IEEE,}    
      Musa~Furkan~Keskin,~\IEEEmembership{Member,~IEEE,}
    Henk~Wymeersch,~\IEEEmembership{Fellow,~IEEE,} and
    Mikko~Valkama,~\IEEEmembership{Fellow,~IEEE}  

\thanks{This work was supported by the ISLANDS project under the EU Horizon Europe (HE) Marie Skłodowska-Curie Actions-Doctoral Networks (MSCA-DN), grant agreement No. 101120544, and by the Swedish Research Council (VR) through the project 6G-PERCEF under Grant 2024-04390.
 }
\thanks{Xi Zhang, Ossi Kaltiokallio, and Mikko Valkama are with the Department of Electrical Engineering, Tampere University, 33100 Tampere, Finland (e-mails: \{xi.2.zhang, ossi.kaltiokallio, mikko.valkama\}@tuni.fi).}
\thanks{Yu Ge, Musa Furkan Keskin, and Henk Wymeersch are with the Department of Electrical Engineering, Chalmers University of Technology, 41296 Gothenburg, Sweden (e-mails: \{yuge, furkan, henkw\}@chalmers.se).}
\vspace{-2mm}
}
\maketitle

\begin{abstract}
\textls[-3]{Radio-based \ac{SLAM} has the potential to provide precise \ac{UE} localization and environmental sensing capabilities by exploiting radio signals. Most existing approaches leverage \ac{LOS} and single-bounce \ac{NLOS} paths solely, while higher-order \ac{NLOS} paths are treated as disturbance. In this paper, we investigate the benefits of leveraging double-bounce \ac{NLOS} paths for solving the bistatic snapshot radio \ac{SLAM} problem. We derive the \ac{CRB} for joint estimation of the \ac{UE} state and landmark positions when double-bounce \ac{NLOS} paths are present. In addition, we propose an algorithm to identify double-bounce \ac{NLOS} paths and leverage them into joint \ac{UE} and landmarks estimation. The derived bounds are validated through simulated data, and the proposed algorithms are evaluated using experimental \ac{mmwave} measurements harnessing beamformed 5G cellular reference signals. The numerical and experimental results demonstrate that the double-bounce \ac{NLOS} paths which share at least one \ac{IP} with the single-bounce \ac{NLOS} paths improve the estimation accuracy of the \ac{UE} state and existing \acp{IP} of single-bounce \ac{NLOS} paths. Importantly, exploiting double-bounce \ac{NLOS} paths enhances environmental mapping capabilities by revealing landmarks that are unobservable with single-bounce \ac{NLOS} paths alone.}  
\end{abstract}
\vspace{-2mm}

%
\IEEEpeerreviewmaketitle

\vspace{-6mm}
\section{Introduction}
\Ac{ISAC} is emerging as one key enabler towards 6G wireless networks, as it jointly offers high data-rate communication and precise sensing capabilities \cite{gonzalez2024integrated}. One promising \ac{ISAC} application is radio-based \acf{SLAM}, which leverages radio signals to localize the \acf{UE} while mapping the surrounding environment \cite{rastorgueva2024millimeter}. In general, various types of radio signals can be utilized for this purpose, including UWB \cite{leitinger2019belief}, 5G \cite{yang2023angle}, and future 6G signals \cite{lotti2023radio}. In addition, different sensing configurations are possible, such as monostatic \cite{kim2023ris, premachandra2023uwb}, bistatic \cite{10818978, kaltiokallio2024integrated}, and multistatic \cite{zhang2019multistatic, tagliaferri2024cooperative}. Among these options, bistatic radio \ac{SLAM} using cellular networks is particularly appealing. It provides full positioning and mapping functionality with only a single \ac{BS}, which allows minimal infrastructure and eliminates the need for synchronization across multiple \acp{BS} \cite{rastorgueva2024millimeter, 10818978, fascista2021downlink, ge2023mmwave}. 

\begin{figure}[!t]
    \centering
    {\includegraphics[trim={1.0cm 5.0cm 1.0cm 5.0cm},width=0.47\textwidth]{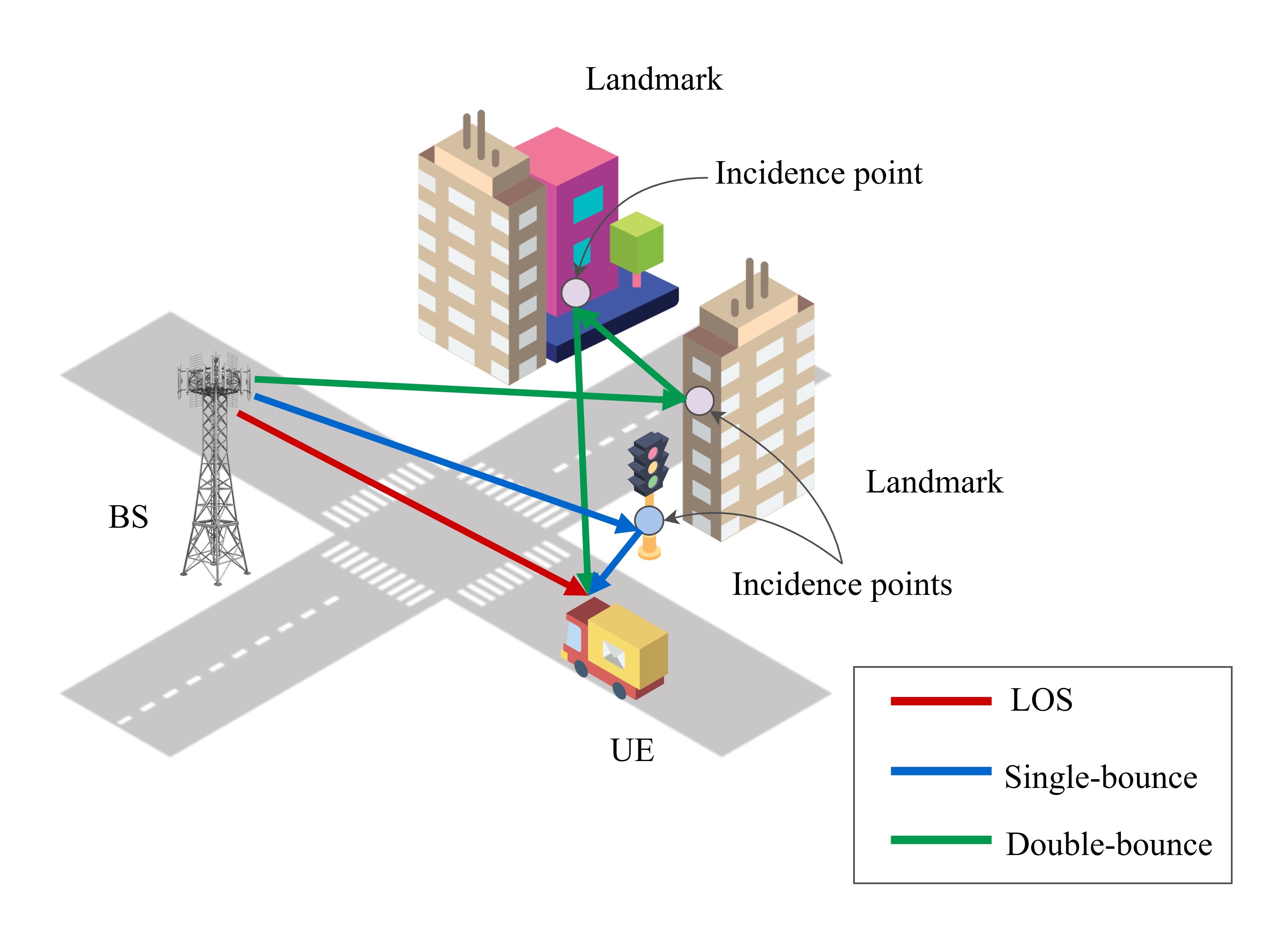}}
    \caption{Illustration of the bistatic radio \ac{SLAM} problem including \ac{LOS}, single-bounce, and double-bounce \ac{NLOS} paths. Double-bounce \ac{NLOS} paths enable improved environmental sensing capabilities since they can be used to illuminate and detect landmarks and other objects that cannot be observed with single-bounce paths alone.}
    \vspace{3mm}
    \label{fig:fig_system model}
\end{figure}

Radio \ac{SLAM} algorithms are commonly categorized into snapshot \cite{10818978, rastorgueva2024millimeter}, filtering \cite{ge2022computationally, kaltiokallio2021mmwave}, and smoothing \cite{ge2025batch,liang2025direct}
based approaches. These categories differ in how they operate over time. Snapshot methods (e.g., \ac{LS} and 
\ac{MLE})\cite{10818978, rastorgueva2024millimeter}  solve each epoch independently based on current measurements, without propagating motion estimates, state information, or any prior knowledge forward in time. Filtering methods (e.g., \ac{EKF} and \ac{PF}) \cite{ge2022computationally, kaltiokallio2021mmwave} perform recursive Bayesian estimation by propagating the filtering density sequentially via time and measurement updates. Smoothing approaches (e.g., GraphSLAM and batch optimization)\cite{ge2025batch,liang2025direct} jointly optimize over a batch of time steps to obtain trajectory estimates. All three approaches rely on signals traveling along the \acf{LOS} path and \acf{NLOS} paths, caused by reflections, scattering, and diffraction, to jointly infer the \ac{UE} state and the positions of environmental landmarks. Most existing works consider only direct \ac{LOS} and single-bounce \ac{NLOS} paths, typically assuming that all the \ac{NLOS} paths are single-bounce paths \cite{ ge20205g, gentner2016multipath, witrisal2016high, kim2022pmbm}. This assumption leads to severe performance degradation when multi-bounce paths are present \cite{10818978, li2023iterative}. Moreover, disregarding multi-bounce paths leaves potential information unexploited, especially in rich multipath environments where higher-order \ac{NLOS} paths can be strong and geometrically structured \cite{ling2017experimental, ju2021millimeter}. An example scenario is illustrated in Fig.~\ref{fig:fig_system model}, where the double-bounce \ac{NLOS} path encode rich environmental maps.

\textls[0]{The fundamental performance limits of \ac{UE} state estimation in the presence of \ac{NLOS} paths have been investigated in \cite{guerra2018single, shahmansoori2017position, mendrzik2018harnessing, 
kakkavas2019performance}. In \cite{guerra2018single}, a single anchor localization scenario was considered and the \ac{FIM} for the \ac{UE} position and orientation, as well as the \ac{NLOS} parameters were presented. Similarly, \cite{shahmansoori2017position} derived the \ac{FIM} for channel parameters and further obtained a closed-form expression for the \ac{FIM} of the \ac{UE} position and orientation by leveraging the geometric relationship between channel parameters and the \ac{UE} position and orientation. Continuing with this idea, \cite{mendrzik2018harnessing} applied a geometric transformation to convert the channel parameter \ac{FIM} into the \ac{FIM} of \ac{UE} position, orientation, and locations of single-bounce \acfp{IP}, and the influence of different \ac{UE}, \ac{BS} and \acp{IP} geometries was also investigated. Likewise, the effect of the Doppler shift on the \ac{FIM} as well as difference in uplink and downlink positioning were considered in  \cite{kakkavas2019performance}. Collectively, the above works show that exploiting \ac{NLOS} components can benefit the \ac{UE} position and orientation estimation. However, all the analyses above are limited to single-bounce \ac{NLOS} components, and the mapping performance is not discussed, as positions of \acp{IP} are often treated as nuisance parameters.}

\textls[0]{Exploiting multi-bounce \ac{NLOS} paths in radio \ac{SLAM} is non-trivial for two reasons. First, it is difficult to identify which measurements are from single- or multi-bounce paths. Second, the multi-bounce \ac{NLOS} paths are much more difficult to handle algorithmically and therefore, many existing solutions restrict their models to single-bounce paths only, ignoring multi-bounce components entirely\cite{ ge20205g, gentner2016multipath, witrisal2016high}. Several works focus on identifying single-bounce paths and mitigating multi-bounce paths \cite{li2023iterative,zhou2025digital,10818978,hu2024multipath}. In \cite{li2023iterative}, an iterative algorithm based on change detection was proposed to detect single-bounce \ac{NLOS} paths, which was then used to estimate the \ac{UE} position and clock bias. In \cite{zhou2025digital}, a \ac{RANSAC}-based outlier
rejection algorithm was utilized to classify \ac{LOS} and single-bounce \ac{NLOS} paths from multi-bounce \ac{NLOS} paths, and the resulting classification was used for \ac{UE} localization. Considering both positioning and mapping, \cite{10818978} proposed a robust snapshot \ac{SLAM} solution that uses a \ac{RANSAC} inspired method to identify the \ac{LOS} and single-bounce \ac{NLOS} paths, thereby isolating higher-order \ac{NLOS} outliers. Differently, \cite{hu2024multipath} leveraged a Hough transform-based line detection method to separate \ac{LOS} and single-bounce \ac{NLOS} paths from higher-order \ac{NLOS} paths, to perform the \ac{SLAM}. These methods improve robustness by isolating multi-bounce paths, but they do not utilize the potential additional information those multi-bounce paths provide.} 

\textls[-2]{By contrast, some recent works have considered incorporating double-bounce \ac{NLOS} paths into mapping with known poses problems and using them to estimate the positions of scattering points \cite{10694028, liu2025graph}. In \cite{10694028}, an initial map and the \ac{UE} state were first obtained using \ac{LOS} and single-bounce paths, after which double-bounce paths were exploited to reveal additional scattering points. In \cite{liu2025graph}, a graph-based method was proposed to model single-bounce and double-bounce paths and these paths were leveraged to iteratively estimate scattering points given known \ac{UE} position. As for \ac{SLAM}, \cite{leitinger2023data, li2025adaptive} introduced their methods based on a multipath \ac{SLAM} framework that modeled each specular surface as a master virtual anchor. In the method, every double-bounce path is treated as two sequential specular reflections involving two distinct master virtual anchors which are then jointly estimated on a factor graph over time as the \ac{UE} moves. However, all \ac{NLOS} paths are modeled as specular reflections, and reflection points lie on assumed planar surfaces. In summary, the influence and potential of versatile multi-bounce \ac{NLOS} components in radio \ac{SLAM} beyond plain specular reflections are not well are not well captured in the existing works.} 

\textls[-2]{In this paper, we investigate the potential benefits of double-bounce \ac{NLOS} paths in the snapshot radio \ac{SLAM} problem. We first derive the \ac{CRB} for the problem to obtain the lower bounds on the estimation error variances for the \ac{UE} position, orientation, clock bias, and \ac{IP} positions in the presence of double-bounce \ac{NLOS} paths. We also present the \ac{FIM} analysis and discuss the identifiability conditions of the problem. Furthermore, we develop a snapshot radio \ac{SLAM} algorithm that exploits \ac{LOS}, single-bounce \ac{NLOS} and double-bounce \ac{NLOS} paths to estimate the unknown \ac{UE} state and \acp{IP} of the single- and double-bounce paths. The devised method does not rely on any prior information of the environmental map nor the \ac{UE} state and it can operate in mixed \ac{LOS}/\ac{NLOS} conditions. Moreover, the method does not assume any specific propagation mechanisms and works for both specular reflection and diffuse scattering, or the combination of them such that a double-bounce path can be reflected by one surface and scattered by another object as it traverses from the transmitter to the receiver. 

The main contributions of this paper can be summarized as follows:}
\begin{itemize}
    \item We derive the fundamental performance bounds for the snapshot radio \ac{SLAM} problem considering \ac{LOS}, single-bounce, and double-bounce \ac{NLOS} paths. The \ac{FIM} analysis reveals that the \ac{SLAM} problem is identifiable under general geometric conditions when a double-bounce \ac{NLOS} path shares at least one of its \acp{IP} with single-bounce \ac{NLOS} paths. The FIM analysis also reveals that the double-bounce \ac{NLOS} paths improve estimation accuracy of the \ac{UE} and \acp{IP} in such conditions.
    \item Inspired by the \ac{FIM} analysis, instead of discarding double-bounce \ac{NLOS} paths and treating them as outliers, we devise a snapshot radio \ac{SLAM} algorithm that exploits them. The proposed method recycles double-bounce \ac{NLOS} paths to enhance \ac{SLAM} performance, turning them from foe to friend.
    \item We validate the proposed algorithm using real-world \ac{mmwave} measurement data, showing that it achieves higher estimation accuracy compared to the baseline methods that discard double-bounce \ac{NLOS} paths. Moreover, the experimental results demonstrate that double-bounce \ac{NLOS} paths enhance environmental mapping capabilities by revealing landmarks that are unobservable with single-bounce NLoS paths alone.
\end{itemize}

The rest of this paper is organized as follows. Section \ref{sec-PF} formulates the problem and describes the fundamentals of the system model. The  performance bounds for the considered problem are derived in Section \ref{sec-Bound} while the proposed snapshot radio \ac{SLAM} algorithm is introduced in Section \ref{sec-estimator}. The numerical and experimental results are presented in Section \ref{sec-results-combined}, and the conclusions are drawn thereafter. Selected technical details are provided in the Appendix.


\textls[-3]{
\emph{Notations:} Scalars are denoted by normal font (e.g., $a$), vectors in lowercase bold letters (e.g., $\va$), and matrices in capital bold letters (e.g., $\vA$). The operations $(\cdot)^\top$, $(\cdot)^{-1}$, $(\cdot)^{\mathrm{H}}$, $|\cdot|$, $\left \| \cdot \right \| $ and $\text{diag}(\cdot)$ denote the transpose, inverse, Hermitian transpose, cardinality of a set (or absolute value when applied to a scalar), Euclidean norm, and the diagonal matrix, respectively. A multivariate Gaussian distribution with mean $\boldsymbol{\mu}$ and covariance $\boldsymbol{\Sigma}$ is denoted as $\mathcal{N}(\boldsymbol{\mu}, \boldsymbol{\Sigma})$.}

\vspace{-3mm}
\section{Problem Formulation \label{sec-PF}}

We consider a bi-static 2D (azimuth domain) downlink scenario with a single \ac{BS} and a single \ac{UE}, where the \ac{BS} acts as the transmitter and the \ac{UE} acts as the receiver. The \ac{BS} has known position $\vp_\textrm{BS} = [x_\textrm{BS}, \, y_\textrm{BS}]^\top$ and orientation $\alpha_\textrm{BS}$. The \ac{UE} state is unknown and given by $\vs = [\vp_\textrm{UE}^\top, \,  \alpha_\textrm{UE}, \, b_\textrm{UE}]^\top \in \mathbb{R}^{4} $, where $\vp_\textrm{UE}$ is the position, $\alpha_\textrm{UE}$ the heading, and $b_{\textrm{UE}}$ the clock bias between the unsynchronized \ac{UE} and \ac{BS} clocks. The $n$th unknown landmark position is represented by $\vm_{n} = [x_n, \, y_n]^\top$ and the $N$ landmarks form a map of the propagation environment. Furthermore, let $\vz_i = [\tau_i, \, \phi_i, \, \theta_i]^\top$ denote the channel parameter estimate of the $i$th propagation path, which consists of the \ac{TOA}, \ac{AOD} and \ac{AOA}. The existence of the \ac{LOS} and the number of resolvable \ac{NLOS} components depends on the problem geometry and communication system parameters, and the set of channel parameter estimates is denoted using $\cZ = \{\vz_0, \vz_1, \ldots, \vz_M \}$ in which $i=0$ is reserved for the \ac{LOS} component and $M$ denotes the number of \ac{NLOS} paths. 

It is to be importantly noted that in general, it is also unknown whether $\vz_i$ is the \ac{LOS}, single-bounce, double-bounce or higher-order \ac{NLOS} path, or clutter due to failure in the channel estimation routine. This paper aims to provide theoretical insights into double-bounce snapshot \ac{SLAM} via \ac{FIM} analysis, as well as, devising a snapshot radio \ac{SLAM} algorithm that jointly estimates the \ac{UE} state and landmark positions using the elements of $\cZ$ that correspond to \ac{LOS}, single-bounce \ac{NLOS} and double-bounce \ac{NLOS}.
  
\vspace{-2mm}
\subsection{Signal Model}
The \ac{BS} transmits \ac{mmwave} downlink reference signals to the \ac{UE} using \ac{OFDM}. A concrete practical example is the \ac{PRS} defined in the 3GPP 5G NR standard. Both the \ac{BS} and \ac{UE} employ antenna arrays with beamforming. Subcarriers are indexed by $k$, \ac{OFDM} symbols by $l$, \ac{BS} transmit beams by $p$, and \ac{UE} receive beams by $q$. We assume further the \ac{OFDM} symbols are captured over a time interval in which the channel is assumed constant. For a multipath radio propagation environment with resolvable path index set $\cI = \{0,1,\ldots,M \}$, the received sample is of the form \cite{rastorgueva2024millimeter}

\begin{equation}
    y_{k,l}^{p,q} = \sum_{i\in\cI} \xi_iG_\textrm{BS}^{p}(\phi_i)G_\textrm{UE}^{q}(\theta_i)e^{-j2\pi k \Delta f \tau_i}x_{k,l}^{p,q} + n_{k,l}^{p,q},
    \label{received signal model}
\end{equation}
where $x_{k,l}^{p,q}$ is the transmitted reference sample, $n_{k,l}^{p,q}$ denotes the additive noise at the \ac{UE} side, and $\Delta f$ is the subcarrier spacing. In the above, the complex path gain, \ac{TOA}, \ac{AOD}, and \ac{AOA} of the $i$th propagation path are $\xi_i$, $\tau_i$, $\phi_i$, and $\theta_i$, respectively, and $i=0$ is reserved for the \ac{LOS} path. Furthermore, $G_\textrm{BS}^{p}(\phi)= \va_{\textrm{BS}}^{\mathrm{H}}(\phi)\vw_p$ and $G_\textrm{UE}^{q}(\theta)=\vw_q^{\mathrm{H}}\va_{\textrm{UE}}(\theta)$ denote the angular responses, including the steering vectors at the \ac{BS} $\va_{\textrm{BS}}(\phi)$ and \ac{UE} $\va_{\textrm{UE}}(\theta)$, and $p$th \ac{BS} beamformer $\vw_p$, $q$th \ac{UE} beamformer $\vw_q$. 

Various channel parameter estimation methods can be used to recover $\cZ$ from \eqref{received signal model} with high accuracy \cite{rastorgueva2024millimeter,shahmansoori2017position, koivisto2021channel}. However, the physical channel parameter estimation is outside the main scope of this paper and it is thus assumed that $\cZ$ is readily available. Since $\cZ$ is used as a measurement input to radio \ac{SLAM}, the channel parameter estimates are referred to as measurements from now on. 

\vspace{-3mm}
\subsection{Geometric Measurement Model}
Let us consider the channel parameter measurement of the $i$th path, $\vz_i = [\tau_i, \, \phi_i, \,\theta_i]^\top$, and let us denote the \acp{IP} of the path using set $\cM_i = \{\vm_i^1, \vm_i^2, \ldots, \vm_i^m \}$ which consists of $m$ physical interactions with the environment ($\cM_0 = \emptyset$ for the \ac{LOS} path). Assuming that the measurement noise is zero-mean Gaussian, which is a common assumption in radio \ac{SLAM} \cite{rastorgueva2024millimeter,kim2022pmbm}, the likelihood function is given by 
\begin{equation}\label{eq:likelihood_function}
    p(\vz_i \mid \vs, \cM_i) = \N(\vz_i; \vh_i(\vs, \cM_i), \vR_i),
\end{equation}
where $\vR_i = \diag([\sigma_\tau^2, \, \sigma_\phi^2, \, \sigma_\theta^2])$ is the covariance matrix and the mean is given by \cite{10818978,10694028}
\begin{equation}\label{eq:geometric_model}
    \vh_i(\vs, \cM_i) = \begin{bmatrix}
        d_i/c + b_\textrm{UE} \\
        \atan2(-\delta_{\phi,i}^{y}, -\delta_{\phi,i}^{x}) - \alpha_\textrm{BS} \\
        \atan2(\delta_{\theta,i}^{y}, \delta_{\theta,i}^{x}) - \alpha_\textrm{UE}
    \end{bmatrix}
\end{equation}
in which $c$ is the speed of light and $\textrm{atan2}(y,x)$ is the two argument arctangent. The mean represents the geometric connection among the \ac{BS}, \ac{UE} and $\cM_i$. The parameters of the model in \eqref{eq:geometric_model} for \ac{LOS}, single-bounce \ac{NLOS} and double-bounce \ac{NLOS} are defined in Appendix \ref{sec:model_parameters_and_jacobian}. Higher-order \ac{NLOS} paths $(m\ge 3)$ are not detailed here as they are not used for estimating the \ac{UE} state and \acp{IP}.

\vspace{-3mm}
\section{Performance Bounds for Double-bounce Snapshot SLAM\label{sec-Bound}}

In this section, we present the performance bounds for the snapshot \ac{SLAM} problem considering \ac{LOS}, single-bounce, and double-bounce paths. Subsequently, we derive the corresponding \ac{FIM} and analyze the impact of incorporating double-bounce paths in the presence of \ac{LOS} and single-bounce paths. 
\vspace{-4mm}
\subsection{Performance Bounds} \label{sec:performance_bounds}

\textls[-3]{The \ac{CRB} states that the \ac{MSE} of an unbiased estimator is always larger than the inverse of the \ac{FIM} in the positive semi-definite sense \cite[Ch. 4.2]{van2004detection}. For notational convenience, let us assume the \ac{LOS} path exists\footnote{This assumption is made for analytical convenience. The derived performance bounds can be extended naturally to \ac{LOS} absent cases and the unknown parameters are identifiable if a sufficient number of single-bounce NLoS paths are available.} and that $d_\vs = \dim(\vs)$, $d_\vm = \dim(\vm_n)$ and $d_\vz = \dim(\vz_i)$ denote the \ac{UE}, \ac{IP} and measurement dimensions, respectively. 
Furthermore, let $\vx = [\vs^\top, \, \vm_1^\top, \, \ldots, \, \vm_N^\top]^\top \in \mathbb{R}^{(d_\vs + N d_\vm) \times 1}$ denote the joint state of the \ac{UE} and $N$ landmarks with dimension $d_\vx = \dim(\vx)$, and $\vz = [\vz_0^\top, \, \vz_1^\top, \, \ldots, \, \vz_M^\top]^\top \in \mathbb{R}^{(M+1) d_{\vz} \times 1}$ denote the concatenated measurement vector. In the following, we assume the order and index of elements in $\cM_i$ are known for every $\vz_i$.\footnote{This assumption is reasonable when deriving the lower bounds on estimation accuracy since we are interested in the best achievable accuracy obtained with perfect data association. We also note that similar assumptions are typically made when computing the \ac{CRB} \cite{wymeersch2018,kaltiokallio2021mmwave}, and the estimator developed in Section \ref{sec-estimator} does not assume the elements in $\cM_i$, nor the order, are known.}  The estimation error \ac{CRB} has the form \cite[Ch. 3.4]{kay1993}}
\begin{equation}\label{eq:cramer_rao_bound}
\mathbb{E}[ \left(\vx - \hat{\vx}(\vz) \right)\left( \vx - \hat{\vx}(\vz) \right)^\top ] \succeq \vJ(\vx)^{-1},
\end{equation}
where $\hat{\vx}(\vz)$ denotes an estimator of $\vx$ and $\vJ(\vx)$ is the $d_\vx \times d_\vx$ \ac{FIM} with elements
\begin{equation}\label{eq:fim_elements}
[\vJ(\vx)]_{i,j}=\mathbb{E}\left\lbrack-\frac{\partial^{2}\log p(\vz \mid \vx)}{\partial {\vx}_{i} \partial {\vx}_{j}} \right\rbrack \qquad i,j=1,\cdots,d_\vx,
\end{equation}
where $p(\vz \mid \vx)$ denotes the joint likelihood function. After the \ac{FIM} is computed as explained above, the \ac{PEB}, \ac{OEB}, \ac{BEB}, \ac{LEB} of $n$th landmark, can be computed as: $\text{PEB} = ([ (\vJ(\vx))^{-1} ]_{1,1} + [ (\vJ(\vx))^{-1} ]_{2,2})^{1/2}$, $\text{HEB} = ([ (\vJ(\vx))^{-1} ]_{3,3})^{1/2}$, $\text{BEB} = ([ (\vJ(\vx))^{-1} ]_{4,4})^{1/2}$, and  $\text{LEB}_n = ({\sum_{i = 2n+3}^{2n + 4} [ (\vJ(\vx))^{-1} ]_{i,i}})^{1/2}$.

\vspace{-2mm}
\subsection{FIM for the Snapshot SLAM}

\begin{figure}[!t]
    \centering
    {\includegraphics[trim={1.0cm 3.1cm 1.0cm 3.5cm},clip,width=\columnwidth]{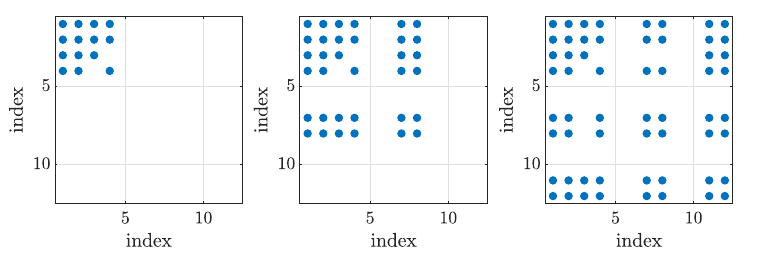}}
    \vspace{-8mm}
    \caption{Visualization of $\vJ_i(\vx)$ in \eqref{eq:fim-per-path} for different path types. \ac{LOS} path visualized on the left ($i \in \cI_\textrm{LoS}, \; \cM_i = \emptyset$), single-bounce path in the middle ($i \in \cI_\textrm{SB}, \; \cM_i = \{\vm_2\}$), and double-bounce path shown on the right ($i \in \cI_\textrm{DB}, \; \cM_i = \{\vm_2, \vm_4\}$). In the example, there are four IPs overall ($N=4$) and the non-zero elements of $\vJ_i(\vx)$ are illustrated using blue markers. 
    }
    \vspace{3mm}
    \label{fig:fim_decomposition}
\end{figure}

The \ac{FIM} defined in \eqref{eq:fim_elements} can be written as \cite[Ch. 3.9]{kay1993}
\begin{equation}\label{eq:fim}
\vJ(\vx) = \vH(\vx)^\top \vR^{-1} \vH(\vx), 
\end{equation} 
where $\vH(\vx) = \partial \vh(\vx)/\partial \vx \in \mathbb{R}^{(M+1)d_{\vz} \times d_\vx}$ is the Jacobian of the concatenated measurement model $\vh(\vx) = [\vh_0(\vs,\cM_0)^\top, \, \vh_1(\vs,\cM_1)^\top, \, \ldots, \, \vh_M(\vs,\cM_M)^\top]^\top \in \mathbb{R}^{(M+1)d_{\vz} \times 1}$ and $\vR = \blkdiag(\vR_0, \, \vR_1, \ldots, \vR_M)$ is the block-diagonal covariance matrix.\footnote{The block-diagonal structure of $\vR$ encodes the assumption that the measurements from different paths are uncorrelated. In practice, correlations may exist, and ignoring them discards some information. This assumption is adopted here to enable a tractable analysis of the considered problem.} Since the measurements are assumed independent and thus the Fisher information is additive over the paths, $\vJ(\vx)$ can be constructed by summing together the \acp{FIM} of the individual paths, expressed as
\begin{equation}
    \vJ(\vx)= \sum_{i \in \cI} \vJ_i(\vx),
\end{equation}
where the \ac{FIM} of the $i$th propagation path is given by
\begin{equation}\label{eq:fim-per-path}
\vJ_i(\vx) = \vH_i(\vx)^\top \vR_i^{-1} \vH_i(\vx),
\end{equation} 
in which $\vH_i(\vx) = \partial \vh_i(\vx)/\partial \vx \in \mathbb{R}^{d_{\vz} \times d_\vx}$. It should be noted that the set of measurement indices is the union of the disjoint \ac{LOS}, single-bounce and double-bounce index sets, that is, $\cI=\cI_\textrm{LoS} \cup \cI_\textrm{SB} \cup \cI_\textrm{DB}$. The structure of $\vJ_i(\vx)$ for different path types is illustrated in Fig. \ref{fig:fim_decomposition} as an example.

\vspace{-2mm}
\subsection{\ac{FIM} for Single-bounce Snapshot \ac{SLAM}} \label{sec:fim_single_bounce_path}
When considering only the \ac{LOS} and single-bounce paths, the \ac{FIM} of the problem can be written as 
\begin{equation}
    \tilde{\vJ}(\vx) = \begin{bmatrix}
        \tilde{\vJ}_{\vs\vs} & \tilde{\vJ}_{\vs\vm}  \\ \tilde{\vJ}_{\vs\vm} ^\top & \tilde{\vJ}_{\vm\vm} 
    \end{bmatrix}.
    \label{eq-fim-los+sb}
\end{equation}
Block $\tilde{\vJ}_{\vs\vs}$ contains the information contributed to the \ac{UE} state, $\tilde{\vJ}_{\vm\vm}$ is the information about the \acp{IP} observed from single-bounce paths, and $\tilde{\vJ}_{\vs\vm}$ captures the coupling between the \ac{UE} and \acp{IP} in the measurement. Let $\ve_i \in \mathbb{N}^{|\cI_\textrm{SB}|\times1}$ denote an all-zeros vector with a 1 in the $i$th element and let $\vE_i = \diag(\ve_i)$ denote an indicator matrix. Now, the sub-blocks in \eqref{eq-fim-los+sb} can be computed as
\begin{align}
    \tilde{\vJ}_{\vs\vs} &= \vH_0(\vs)^\top \vR_0^{-1} \vH_0(\vs) + \sum_{i\in \mathcal{I}_{\textrm{SB}}} \vH_i(\vs)^\top \vR_i^{-1} \vH_i(\vs),\\
    \tilde{\vJ}_{\vs\vm} &= \sum_{i\in \mathcal{I}_{\textrm{SB}}} \left(\ve_i^\top\otimes\left( \vH_i\left(\vs\right)^\top \vR_i^{-1} \vH_i\left(\vm_i\right)\right)\right),\\
    \tilde{\vJ}_{\vm\vm} &= \sum_{i\in \mathcal{I}_{\textrm{SB}}} \left( \vE_i \otimes \left(\vH_i\left(\vm_i\right)^\top \vR_i^{-1} \vH_i\left(\vm_i\right)\right)\right),
\end{align}
where $\vH_0(\vs)$ is the Jacobian of \ac{LOS} measurement with respect to the \ac{UE} state given in \eqref{eq:jacobian_los_ue}, $\vH_i(\vs)$ and $\vH_i(\vm_i)$ are the Jacobians of $i$th single-bounce measurement with respect to the \ac{UE} state and its associated \ac{IP}, given in \eqref{eq:jacobian_sb_ue} and \eqref{eq:jacobian_sb_xl}, respectively.

\begin{figure}[!t]
    \centering
    \includegraphics[width=0.9\columnwidth]{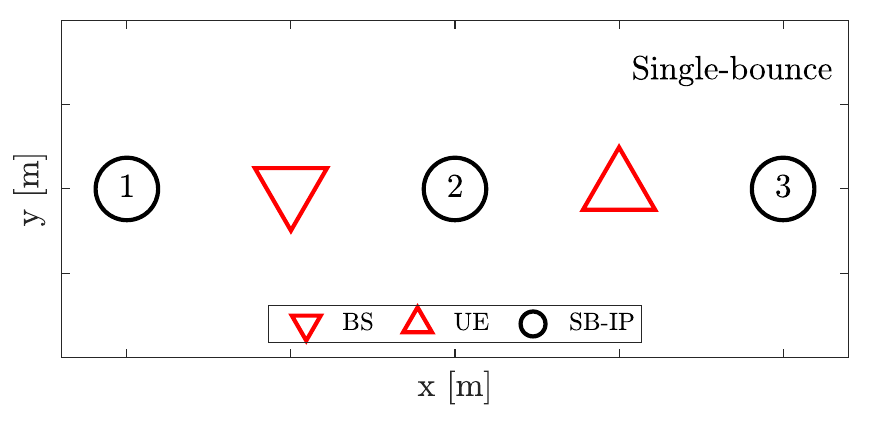}
    \vspace{-2mm}
    \caption{Examples of rank-deficient single-bounce SLAM geometries. In the examples, the \ac{LOS} is assumed to exist and one single-bounce \ac{IP} is considered such that $\dim(\vx) = 6$. In scenario 1, as indicated by the number inside the marker, the rank of the \ac{FIM} is $\textrm{rank}(\vJ(\vx)) = 5$ and null space of the system is $\textrm{null}(\vH(\vx)) = \begin{psmallmatrix} -1 & 0 & 0 & 1 & 0 & 0 \end{psmallmatrix}^\top$ meaning that we can change the $x$ coordinate of the UE and the clock bias (in equal but opposite values) without changing the measurements. In scenario 3, $\textrm{rank}(\vJ(\vx)) = 5$ and  $\textrm{null}(\vH(\vx)) = \begin{psmallmatrix} 1 & 0 & 0 & -1 & 1 & 0 \end{psmallmatrix}^\top$, whereas in scenario 2, $\textrm{rank}(\vJ(\vx)) = 4$ and $\textrm{null}(\vH(\vx)) = \begin{psmallmatrix} -1 & 0 & 0 & 1 & 0 & 0 \\ 0 & 0 & 0 & 0 & 1 & 0 \end{psmallmatrix}^\top$.}
    \vspace{4mm}
    \label{fig:observability_analysis}
\end{figure}

Note that the \ac{LOS} path contributes only to $\tilde{\vJ}_{\vs\vs}$, whereas each single-bounce path contributes to $\tilde{\vJ}_{\vs\vs}$, $\tilde{\vJ}_{\vm\vm}$, and the cross-information $\tilde{\vJ}_{\vs\vm}$ (linking $\vs$ and the corresponding $\vm_n$). By analyzing the \ac{FIM} in \eqref{eq-fim-los+sb}, several studies have demonstrated that unambiguous \ac{UE} state estimation is feasible if a sufficient number of single-bounce \ac{NLOS} paths are available, even if the \ac{LOS} path is absent \cite{guerra2018single, shahmansoori2017position, mendrzik2018harnessing, 
kakkavas2019performance}. Furthermore, the single-bounce \ac{NLOS} paths can provide \ac{UE} position and orientation information, thereby improving the accuracy of \ac{UE} state estimation. For the \ac{SLAM} problem considered in this work and as shown in \cite{10818978}, the problem is solvable with either (i) the \ac{LOS} and one single-bounce \ac{NLOS}, or (ii) four single-bounce \ac{NLOS}. However, certain geometric configurations can lead to a rank-deficient \ac{FIM}. Examples of such degenerate geometries are illustrated in Fig.~\ref{fig:observability_analysis}, with their corresponding rank and null spaces summarized in the caption. As shown, when the \ac{UE}, \ac{BS}, and \ac{IP} are collinear, some parameters become unobservable as indicated by $\textrm{null}(\vH(\vx))$. 

\vspace{-3mm}
\subsection{Impacts of the Double-bounce Path \label{sec:fim_double_bounce_path}}
Now we analyze how the inclusion of a double-bounce path affects the \ac{FIM} and the observability of the problem. Let us assume that using existing \ac{LOS} and single-bounce paths, all unknown parameters can be estimated (i.e., $\tilde{\vJ}(\vx) \succ \boldsymbol{0}$). We then incorporate one additional double-bounce path. The contribution from this path to the \ac{FIM} can be written as 
\begin{equation}
    \vJ_{\text{DB}} = \begin{bmatrix}
        \vA & \vB \\\vB^\top & \vC
    \end{bmatrix},
    \label{eq-double-bounce FIM}
\end{equation}
where the matrix $\vA$ corresponds to the information that the double-bounce path provides about the parameters already involved in the \ac{LOS} and single-bounce paths, $\vC$ is the information about any additional parameters introduced by the double-bounce path (i.e., \acp{IP} that are only observed from the double-bounce path), and $\vB$ captures the cross-information between the existing parameters and the parameters introduced solely from the double-bounce path. Consequently, the overall \ac{FIM} combining all paths can be formed as 
\begin{equation}
    \vJ(\vx) = \begin{bmatrix}
        \tilde{\vJ}(\vx) + \vA & \vB \\\vB^\top & \vC
    \end{bmatrix}. 
    \label{eq-overall FIM}
\end{equation}
 The structure of matrices $\vA$, $\vB$, and $\vC$ depends on the geometric relationship between the \acp{IP} of the double-bounce path and those already observed via single-bounce paths. 
%
%
To characterize the resulting information gain on the \ac{UE} state and the existing \acp{IP} and observability properties of the overall \ac{SLAM} problem, we consider three different cases in the following.

\subsubsection{Case 1: Double-bounce path shares both \acp{IP} with single-bounce paths}
In this scenario, both \acp{IP} of the double-bounce path have also been observed through single-bounce paths. Therefore, the double-bounce measurement does not introduce any new unknown parameters. In terms of the \ac{FIM}, this implies $\vB = \emptyset$ and $\vC = \emptyset$, and $\vA$ can be computed using \eqref{eq:fim-per-path} in which the Jacobian is $\vH_i(\vx) = \partial \vh_i(\vx)/\partial \vx \in \mathbb{R}^{d_{\vz} \times d_\vx}$. As an example, the Jacobian for the double-bounce path considered in Fig.~\ref{fig:fim_decomposition} is given by
\begin{equation}\label{eq:jacobian_case1}
    \vH_i(\vx) = \begin{bmatrix}
        \vH_i(\vs) & \mathbf{0}_{d_\vz \times d_\vm} & \vH_i(\vm_2) & \mathbf{0}_{d_\vz \times d_\vm} & \vH_i(\vm_4) 
    \end{bmatrix},
\end{equation}
where $\vH_i(\vs)$, $\vH_i(\vm_2)$ and $\vH_i(\vm_4)$ are given in equations \eqref{eq:jacobian_db_ue}, \eqref{eq:jacobian_db_xl1} and \eqref{eq:jacobian_db_xl2}, respectively. As illustrated in Fig.~\ref{fig:fim_decomposition}, $\vA$ contains non-zero entries only in rows/columns corresponding to the \ac{UE} state and the two involved \acp{IP}. 
Since $\vA \succeq \mathbf{0}$ and $\vJ(\vx) =\tilde{\vJ}(\vx) + \vA \succeq \tilde{\vJ}(\vx) \succ \mathbf{0}$, we have $(\vJ(\vx))^{-1} \preceq (\tilde{\vJ}(\vx))^{-1}$, and thus, for every parameter the corresponding $\ac{CRB}$ cannot increase. 

To summarize, the inclusion of such double-bounce path directly improves the estimation accuracy of the \ac{UE} state and the two associated \acp{IP}. Moreover, it can indirectly improve the estimation accuracy of other single-bounce \acp{IP} through their coupling with the \ac{UE} state. Consequently, the covariance matrix of all estimated parameters is reduced, even for parameters that are not directly involved in the double-bounce path.

\subsubsection{Case 2: Double-bounce path shares one \ac{IP} with one of the single-bounce paths}
In this case, one \ac{IP} of the double-bounce path is seen by an existing single-bounce path, and another \ac{IP} is added which has not been observed before. Let us denote the existing state vector as $\tilde{\vx} = [\vs^\top, \vm_1^\top, \cdots, \vm_{N-1}^\top]^\top$, and assume the double-bounce path involves one \ac{IP} $\vm_{N-1}$ and one new \ac{IP} $\vm_{N}$. In \eqref{eq-double-bounce FIM}, $\vA$, $\vB$ and $\vC$ are $(d_\vs+(N-1)d_\vm) \times(d_\vs+(N-1)d_\vm)$, $(d_\vs+(N-1)d_\vm) \times d_\vm$, $d_\vm \times d_\vm$ non-zero block matrices, respectively. 

To understand the effect of this double-bounce path on estimating the existing parameters $\tilde{\vx}$, we consider the \ac{EFIM} of $\tilde{\vx}$, which is given as the Schur complement of $\vC$ in \eqref{eq-overall FIM}
\begin{equation}
    \vJ^e = \tilde{\vJ}(\tilde{\vx}) + \vA - \vB\vC^{-1}\vB^\top,
    \label{eq- EFIM}
\end{equation}
when $\vC$ is invertible. Here, $\vJ^e$ is composed of three terms. The first term $\tilde{\vJ}(\tilde{\vx})$ quantifies the information from \ac{LOS} and single-bounce paths. The second term $\vA$ quantifies the information gain from the double-bounce path if $\vm_N$ were known. And, the third term $\vB\vC^{-1}\vB^\top$ quantifies the loss of information considering the uncertainty of $\vm_N$. 

Since $\vJ_\text{DB}$ is the information matrix meaning that $\vJ_\text{DB} \succeq \mathbf{0}$. When $\vC$ is non-singular, we have $\vC \succ \mathbf{0}$. Therefore, the Schur complement of $\vC$ satisfies that 
\begin{equation}
    \vA - \vB\vC^{-1}\vB^\top \succeq \boldsymbol{0},
    \label{eq-remark1}
\end{equation}
meaning that the net information gain from the double-bounce path to the existing parameters $\tilde{\vx}$ is always non-negative. In other words, adding this double-bounce path cannot degrade the estimation performance of $\tilde{\vx}$.



Next, we give the structure of $\vA$, $\vB$ and $\vC$, to provide a structure of the term $\vA -\vB\vC^{-1}\vB^\top$. Let $i$ be the index of the double-bounce path under consideration. By definition, we can write $\vA$ as 
\begin{equation}
    \vA = \vH_i({\tilde{\vx}})^\top \vR_i^{-1} \vH_i({\tilde{\vx}}),
    \label{eq-alpha}
\end{equation}
and the second term in \eqref{eq-remark1} can be expressed as 
\begin{multline}\label{eq-betainvRbeta}   \vB\vC^{-1}\vB^\top = \vH_i\left( \tilde{\vx} \right)^\top \vR_i^{-1}\vH_i\left(\vm_{N}\right) \\
 \times \left(\vH_i\left(\vm_N\right)^\top \vR_i^{-1}\vH_i\left(\vm_N\right)\right)^{-1}   \vH_i\left( \vm_N \right)^\top \vR_i^{-1} \vH_i\left( \tilde{\vx} \right), 
\end{multline}
where $\vH_i\left( \tilde{\vx} \right)$ is the Jacobian of the double-bounce measurement with respect to $\tilde{\vx}$, which can be expressed as $\vH_i\left( \tilde{\vx} \right) = [\vH_i({\vs}),\mathbf{0}_{d_\vz \times (N-1)d_\vm},\vH_i({\vm_{N-1}})]$. The Jacobians $\vH_i({\vs})$, $\vH_i({\vm_{N-1}})$, and $\vH_i\left( \vm_N \right)$ are given in \eqref{eq:jacobian_db_ue}, \eqref{eq:jacobian_db_xl1} and \eqref{eq:jacobian_db_xl2}, respectively. For notational simplicity, let $\vH_{i,\tilde{\vx}}\triangleq \vH_i\left( \tilde{\vx} \right)$, $\vH_{i,\vm} \triangleq\vH_i\left(\vm_N \right)$. We can perform Cholesky decomposition of $\vR_i^{-1}$ such that $\vR_i^{-1} = \vW^{\top}\vW$ and define $\tilde{\vH}_{i,\tilde{\vx}} = \vW\vH_{i,\tilde{\vx}}$ and $\tilde{\vH}_{i,\vm} = \vW\vH_{i,\vm}$. Substituting these into \eqref{eq-alpha} and \eqref{eq-betainvRbeta} gives
\begin{align}
\vA &= \tilde{\vH}_{i,\tilde{\vx}}^\top\tilde{\vH}_{i,\tilde{\vx}}, \\
\vB\vC^{-1}\vB^\top& = \tilde{\vH}_{i,\tilde{\vx}}^\top \tilde{\vH}_{i,\vm}\left( \tilde{\vH}_{i,\vm}^\top\tilde{\vH}_{i,\vm} \right)^{-1}\tilde{\vH}_{i,\vm}^\top \tilde{\vH}_{i,\tilde{\vx}}.
\end{align}
Define next $\vP = \tilde{\vH}_{i,\vm}\left( \tilde{\vH}_{i,\vm}^\top\tilde{\vH}_{i,\vm} \right)^{-1}\tilde{\vH}_{i,\vm}^\top$, which is the orthogonal projector onto the column space of $\tilde{\vH}_{i,\vm}$. Therefore, the difference of interest can be written as 
\begin{equation}
\vA - \vB\vC^{-1}\vB^\top = 
\tilde{\vH}_{i,\tilde{\vx}}^\top(\vI-\vP)\tilde{\vH}_{i,\tilde{\vx}}. 
\label{eq-proof-inequality}
\end{equation}
Since the matrix $\tilde{\vH}_{i,\vm}$ has full column rank $d_{\vm}=2$, the projection matrix $\vP$ is also rank two. Consequently, in \eqref{eq-proof-inequality}, the matrix $\vI -\vP$ is rank one, and the matrix $\tilde{\vH}_{i,\tilde{\vx}}^\top(\vI-\vP)\tilde{\vH}_{i,\tilde{\vx}}$ is rank one. 
Therefore, the double-bounce path only provides information in one dimension to the existing parameters $\tilde{\vx}$.
Intuitively, when $\vm_N$ is the last \ac{IP} before reaching \ac{UE}, as shown in \eqref{eq:jacobian_db_xl2}, the Jacobian entries corresponding to the \ac{AOD} measurement are zero, meaning that $\vm_N$ does not influence the \ac{AOD}, and the \ac{AOD} directly contributes to $\tilde{\vx}$.

\begin{figure}[!t]
    \centering
    \includegraphics[width=0.9\columnwidth]{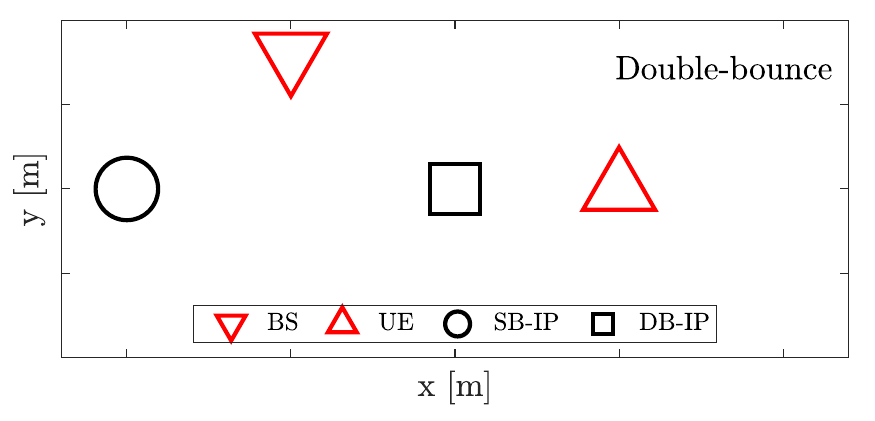}
    \vspace{-2mm}
    \caption{An example of a rank-deficient double-bounce SLAM geometry. The example considers the \ac{LOS}, one single-bounce \ac{NLOS} and one double-bounce \ac{NLOS} that shares one of the \acp{IP} with the single-bounce path. The dimension of the joint state is $\dim(\vx) = 8$, the rank of the \ac{FIM} is $\textrm{rank}(\vJ(\vx)) = 7$ and null space of the system is $\textrm{null}(\vH(\vx)) = \begin{psmallmatrix} 0 & 0 & 0 & 0 & 0 & 0 & 1 & 0 \end{psmallmatrix}^\top$.}
    \vspace{3mm}
    \label{fig:observability_analysis_db}
\end{figure}

It is important to note that the above conclusion holds only if the matrix $\vC$ is invertible (full rank). In certain geometric configurations, $\vH_{i}(\vm_N)$ may not have full column rank, causing $\vC$ to be singular. One such degenerate scenario happens when the $\vm_{N-1}$, $\vm_N$, and the \ac{UE} are collinear and the $\vm_N$ lies between $\vm_{N-1}$ and the \ac{UE}. This is reflected in the Jacobian $\vH_{i}(\vm_N)$: under this collinear geometry, the elements in the first row of $\vH_{i}(\vm_N)$ become zeros (see the expression in \eqref{eq:jacobian_db_xl2}). An example of this rank-deficient geometry is illustrated in Fig.~\ref{fig:observability_analysis_db}, and the resulting \ac{FIM} rank and null space are detailed in the caption. As shown, moving the second \ac{IP} (black square) along the $x$-axis has no effect on the observed measurements, resulting in the $x$-direction of the second \ac{IP} becoming unidentifiable.


\subsubsection{Case 3: Double-bounce path does not share \acp{IP} with any single-bounce paths}
In the third case, both \acp{IP} of the double-bounce path are new and have not been observed from any single-bounce paths. Thus, such path introduces two new unknown vectors into the estimation problem. The \ac{FIM} contribution from the double-bounce in this case has the structure $\vA\in\mathbb{R}^{(d_\vs+(N-2)d_\vm) \times (d_\vs+(N-2)d_\vm)}$, $\vB\in\mathbb{R}^{(d_\vs+(N-2)d_\vm) \times 2d_\vm}$ and  $\vC\in \mathbb{R}^{2d_\vm \times 2d_\vm}$. Importantly, the rank of $\vC \le d_\vz=3$ is always smaller than the dimension of new involved parameters $2d_\vm =4$, and at least one degree of freedom remains unobservable. Thus, $\vC$ is rank-deficient, and the resulting overall \ac{FIM} $\vJ(\vx)$ is also rank-deficient. As a result, if both \acp{IP} of a double-bounce path have not been unobserved elsewhere, adding this double-bounce path does not provide information gain on the \ac{UE} and single-bounce \acp{IP}. Identifying and discarding such double-bounce path, the \ac{SLAM} problem can still be solvable.

In summary, under general geometric conditions, adding double-bounce paths that share at least one \ac{IP} with single-bounce paths results in a solvable \ac{SLAM} problem. Moreover, these double-bounce paths can provide information gain on the \ac{UE} state and all single-bounce \acp{IP}, thus improving the estimation accuracy of these parameters. An exceptional rank-deficient geometry case is illustrated in Fig.~\ref{fig:observability_analysis_db}, where adding a double-bounce that shares one \ac{IP} with a single-bounce path renders the problem unsolvable. In this degenerate case, the double-bounce path can be identified and removed after which the problem remains solvable.

\vspace{-2mm}
\section{Proposed Double-bounce Snapshot\\\ac{SLAM} Algorithms \label{sec-estimator}}
In this section, we describe the proposed method to solve the snapshot \ac{SLAM} problem utilizing measurements from the \ac{LOS}, single-bounce, and double-bounce paths. 

\vspace{-2mm}
\subsection{Proposed Method -- High-Level Approach}
Recall that $\cI=\{0,1,...,M\}$ is the measurement index set which can be partitioned into four disjoint sets as $\cI=\{\cI_\textrm{LoS},\cI_\textrm{SB},\cI_\textrm{DB},\cI_\textrm{HO}\}$, where $\cI_\textrm{LoS}$ is the \ac{LOS} index, $\cI_\textrm{SB}$ and $\cI_\textrm{DB}$  contain the single- and double-bounce indices, respectively, and $\cI_\textrm{HO}$ is the set of indices corresponding to higher-order interactions and clutter measurements, which do not provide any useful information for solving the problem. One of the main difficulties in solving the snapshot \ac{SLAM} problem is correctly classifying the elements in the disjoint sets. To overcome this challenge, we propose a four-step estimator which can be summarized as follows:
\begin{enumerate}
    \item \textit{Initial SLAM solution} -- To compute the initial \ac{UE} and landmark estimates, we resort to the robust snapshot \ac{SLAM} algorithm described in \cite{10818978}. The method partitions the measurements into a set of inliers $\cI_\textrm{inlier} = \{\cI_\textrm{LoS},\cI_\textrm{SB}\}$  which are used to solve the problem, and outliers $\cI_\textrm{outlier} = \{\cI_\textrm{DB},\cI_\textrm{HO}\}$ which are considered as unwanted propagation paths that degrade \ac{SLAM} performance. The method is summarized in Section \ref{sec-snapshot}.
    \item \textit{Double-bounce classification} -- From the set of outliers $i \in \cI_\textrm{outlier}$ obtained in the first step, propagation path $i$ is classified as a double-bounce path based on geometrical consistency with respect to $i' \in \cI_\textrm{SB}$. The classification details of $\cI_\textrm{DB}$ are provided in Section \ref{sec-db-identify}. 
    \item \textit{Estimating \acp{IP} of} $\cI_\textrm{DB}$ -- The identified double-bounce paths share either one or both \acp{IP} with the single-bounce paths. If path $i \in \cI_\textrm{DB}$ only shares one of the \acp{IP} with path $i' \in \cI_\textrm{SB}$, the second \ac{IP} of the double-bounce path can be estimated from the intersection point of a half-line and an ellipse. In Section \ref{sec-closeform2IP}, a closed-form solution for the second \ac{IP} is derived.
    \item \textit{Double-bounce snapshot \ac{SLAM}} -- The \ac{MLE} for the double-bounce snapshot \ac{SLAM} problem is presented in Section \ref{sec:db_mle}. The devised method utilizes $\{\cI_\textrm{LoS},\cI_\textrm{SB},\cI_\textrm{DB}\}$, as well as the initial \ac{UE} and landmark estimates obtained in steps $1-3$.
\end{enumerate}

\vspace{-2mm}
\subsection{Initial SLAM Solution\label{sec-snapshot}}

To solve the \ac{SLAM} problem in mixed \ac{LOS}/\ac{NLOS} conditions, we resort to the robust snapshot \ac{SLAM} algorithm introduced initially in \cite{10818978}. The algorithm uses a \ac{RANSAC} inspired \ac{LS} approach in which the method first uses a minimal subset of the channel parameter estimates $\cZ = \{\vz_0, \vz_1, \ldots, \vz_M \}$ to compute an initial solution $\hat{\vs}_0$. Then, based on $\hat{\vs}_0$ and an error metric, the measurement indices are partitioned into a set of inliers $\cI_\textrm{inlier} = \{\cI_\textrm{LoS},\cI_\textrm{SB}\}$ and outliers $\cI_\textrm{outlier} = \{\cI_\textrm{DB},\cI_\textrm{HO}\}$ such that $\cI = \{\cI_\textrm{inlier}, \cI_\textrm{outlier} \}$, where $\cI_\textrm{inlier} \cap \cI_\textrm{outlier} = \emptyset$. 
Thereafter, the problem is re-solved using $\cI_\textrm{inlier}$ and the \ac{UE} estimate $\hat{\vs}$ is the one that minimizes the \ac{LS} cost. Lastly, the map estimate is given by $\hat{\cM} = \{\hat{\vm}_i \mid i \in \cI_\textrm{SB}\}$ in which the landmark estimates $\hat{\vm}_i$ are computed independently for every $i \in \cI_\textrm{SB}$ by solving a nonlinear optimization problem.

\subsection{Double-bounce Classification \label{sec-db-identify}}

Given the single-bounce index set $\cI_{\textrm{SB}}$ identified in Section \ref{sec-snapshot}, we identify the double-bounce paths $\cI_{\textrm{DB}} \subseteq \cI_{\textrm{outlier}}$ by exploiting angular consistency. Specifically, if a candidate double-bounce path and a single-bounce path share the same \ac{IP} on the transmitter side, their \acp{AOD} are expected to be closely aligned. Or vice versa, if the \ac{IP} is on the receiver side, their \acp{AOA} are expected to be similar.

Let us denote the angular distances between the \acp{AOD} and \acp{AOA} of the candidate double-bounce path $i \in \cI_{\textrm{outlier}}$ and identified single-bounce paths $\cI_{\textrm{SB}}$ using $\delta \vphi_i = [\Delta(\phi_i,\phi_1), \ldots, \Delta(\phi_i,\phi_{\lvert \cI_{\textrm{SB}}\rvert})]$ and $\delta \vtheta_i = [\Delta(\theta_i,\theta_1), \ldots, \Delta(\theta_i,\theta_{\lvert \cI_{\textrm{SB}}\rvert})]$, respectively, in which the angular distance between two angles is defined as 
\begin{equation}
    \Delta(a,b) = \lvert (a - b + \pi ) \Mod{2\pi}   - \pi \rvert.
\end{equation}
Let $\vpsi_i = [\delta \vphi_i, \, \delta \vtheta_i]$ denote the combined angular distance vector. The index with the smallest angle error is given by 
\begin{equation}\label{eq:smallest_angle_error_index}
    n_{\vpsi} = \underset{n}{\argmin} \vpsi_i, \quad n=1,2,\ldots, 2 \lvert \cI_\textrm{SB} \rvert
\end{equation}
and if $[ \vpsi_i]_{n_{\vpsi}}$ is below angular threshold $T_{\vpsi}$, path $i$ is classified as a double-bounce path. Mathematically, $\cI_\textrm{DB}$ is formed as follows
\begin{equation}\label{eq:db_set_builder}
    \cI_\textrm{DB} = \{i \in \cI_{\textrm{outlier}} \mid  [ \vpsi_i]_{n_{\vpsi}} \leq T_{\vpsi} \}. 
\end{equation}

\subsection{Estimating \acp{IP} of $\cI_\text{DB}$ \label{sec-closeform2IP}}

The \acp{IP} $\cM_i = \{\vm_i^1, \vm_i^2 \}$ for double-bounce paths $i \in \cI_\textrm{DB}$ are formed by considering two cases:
\begin{enumerate}
    \item \textit{Case 1:} We use \acp{IP} of two different single-bounce paths as $\cM_i = \{\vm_i^1, \vm_i^2 \}$ if the following logical condition 
    \begin{equation}\label{eq:DB_condition}
        [\delta \vphi_i ]_{n_{\vphi}} \leq T_{\vpsi} \text{ and } [\delta \vtheta_i ]_{n_{\vtheta}} \leq T_{\vpsi},
    \end{equation}
    is true. In the above, the \ac{AOD} and \ac{AOA} indices are given by, $n_{\vphi} = \argmin_n \delta \vphi_i$ and $n_{\vtheta} = \argmin_n \delta \vtheta_i$, respectively. The \acp{IP} of the double-bounce path are $\vm_i^1 = \hat{\vm}_{n_{\vphi}}$ and $\vm_i^2 = \hat{\vm}_{n_{\vtheta}}$.
    \item \textit{Case 2:} We use an \ac{IP} of a single-bounce path as $\vm_i^1$ and the details to estimate $\vm_i^2$ are explained below. If $n_{\vpsi} \leq \lvert \cI_{\textrm{SB}} \rvert$, the \acp{AOD} of the single- and double-bounce paths are similar and we set $\vm_i^1 = \hat{\vm}_{n_{\vpsi}}$. Conversely, if $n_{\vpsi}  > \lvert \cI_{\textrm{SB}} \rvert$, the \acp{AOA} are similar and we set $\vm_i^1 = \hat{\vm}_{n_{\vpsi}  - {\lvert \cI_{\textrm{SB}}\rvert}}$.
\end{enumerate}

Next, we describe how $\vm_i^2$ is estimated in \textit{Case 2}. The method is inspired by \cite{10694028} but, differently, a closed-form derivation is given below. Let us denote the starting point, two \acp{IP} and the end point of $i \in \cI_\textrm{DB}$ using $\cP_i = \{\vp_i^0, \, \vm_i^1, \, \vm_i^2, \, \vp_i^3\}$. If the index with the smallest angle error is $n_{\vpsi} \leq \lvert \cI_\textrm{SB} \rvert$ (see Eq. \eqref{eq:smallest_angle_error_index}), \ac{BS} is considered the starting point ($\vp_i^0 = \vp_\textrm{BS}$) and \ac{UE} the end point ($\vp_i^3 = \hat{\vp}_\textrm{UE}$). Or vice versa if $n_{\vpsi} > \lvert \cI_\textrm{SB} \rvert$, \ac{UE} is considered the starting point ($\vp_i^0 = \hat{\vp}_\textrm{UE}$) and \ac{BS} the end point ($\vp_i^3 = \vp_\textrm{BS}$). The second \ac{IP} is then located on a half-line 
\begin{equation}
    \vm_i^2 = \vp_i^3  + \lambda\vt_i, \quad \lambda > 0,
    \label{half-line}
\end{equation}
that spans from $\vp_i^3$ toward the unit direction vector $\vt_{i}$, given by
\begin{equation}
    \vt_{i} = \begin{cases} 
    \vOmega(\hat{\alpha}_{\text{UE}}) [\cos(\theta_i),\, \sin(\theta_i)]^\top & \text{if} \quad  n_{\vpsi}  \leq |\cI_{\textrm{SB}}|, \\
    \vOmega(\alpha_{\text{BS}}) [\cos(\phi_i),\,\sin(\phi_i)]^\top & \text{otherwise},
    \end{cases}
\end{equation}
in which $\vOmega(\alpha) = \Bigl[\begin{smallmatrix}
    \cos(\alpha) & - \sin(\alpha) \\ \sin(\alpha) & \cos(\alpha)
\end{smallmatrix} \Bigr]$ is a counterclockwise rotation matrix. The second \ac{IP} also resides on an ellipse defined as
\begin{equation}\label{ellipse}
    \lVert  \vm_i^2 -  \vm_i^1 \rVert + \lVert \vm_i^2 - \vp_i^3 \rVert = d_i^e,
\end{equation}
with $d_i^e = c(\tau_i - \hat{b}_{\textrm{UE}}) - \lVert \vp_i^0 -  \vm_i^1 \rVert$ and two foci at $\vm_i^1$ and $\vp_i^3$. Thus, $\vm_i^2$ is at the intersection point of a half-line and an ellipse, and a closed-from solution can be obtained by combining \eqref{half-line} and \eqref{ellipse} and solving for $\vm_i^2$. This is achieved by first plugging \eqref{half-line} into \eqref{ellipse}, which gives
\begin{equation}
    \lVert \vp_i^3  + \lambda\vt_i - \vm_i^1 \rVert + \lVert \vp_i^3 + \lambda\vt_i - \vp_i^3  \rVert =  d_i^e.
    \label{eq-calculate}
\end{equation}
By expanding the Euclidean norms and noting that $\lVert \vp_i^3 + \lambda\vt_i - \vp_i^3 \rVert = \lambda$ since $\vt_i$ is a unit vector, yields
\begin{equation}\label{eq-simplify}
    \sqrt{\lVert \vp_i^3 - \vm_i^1 \rVert^2 + 2 \lambda( \vp_i^3 - \vm_i^1)^\top \vt_i + \lambda^2} = d_i^e - \lambda.
\end{equation}
Taking the square on both sides and solving for $\lambda$ gives
\begin{equation}
    \lambda = \frac{(d_i^e)^2 - \lVert \vp_i^3 - \vm_i^1 \rVert^2}{2((\vp_i^3 - \vm_i^1)^\top \vt_i + d_i^e)},
\end{equation}
and substituting $\lambda$ back into \eqref{half-line} yields the closed-form solution
\begin{equation}
    \hat{\vm}_i^2 = \vp_i^3 + \frac{(d_i^e)^2 - \lVert \vp_i^3 - \vm_i^1 \rVert^2}{2((\vp_i^3 - \vm_i^1)^\top \vt_i + d_i^e)} \vt_i.
\end{equation}
After solving $\vm_i^2$, the feasibility of the solution is checked using $d_i^e > \lVert \vp_i^3 - \vm_i^1  \rVert$. If the solution is feasible, the \ac{IP} is appended to the map, as $\hat{\cM} = \hat{\cM} \cup \hat{\vm}_i^2$, and if the check fails, the candidate path $i$ is removed from $\cI_{\textrm{DB}}$, that is, $\cI_{\textrm{DB}} = \cI_{\textrm{DB}} \setminus i$.

\subsection{Double-bounce Snapshot \ac{SLAM}}\label{sec:db_mle}

Recall that the joint state of the \ac{UE} and \acp{IP} is $\vx = [\vs^\top, \, \vm_1^\top, \, \ldots, \, \vm_{\lvert \cM \rvert}^\top]^\top \in \mathbb{R}^{(d_\vs + \lvert \cM \rvert d_\vm) \times 1}$ in which $\vm_n \in \cM$. The considered double-bounce snapshot \ac{SLAM} problem can be solved by maximizing the joint likelihood function with respect to $\vx$, which is equivalent to minimizing the negative log-likelihood, resulting in the following problem
\begin{equation}\label{eq-loglike}
    \hat{\vx} = \arg\min_{\vx}L(\vx).
\end{equation}
In the above, $L(\vx)$ is the quadratic cost function we wish to minimize and is given by
\begin{equation}
    L(\vx) = \left(\vz - \vh(\vx) \right)^\top \vR^{-1}\left(\vz - \vh(\vx) \right). 
\end{equation}
Since the measurements are independent, we can express the cost function as
\begin{equation}\label{eq-objfunc}
     L(\vx) =  \sum_{i \in \cI_\textrm{inlier}} \left(\vz_i - \vh_i(\vx) \right)^\top \vR_i^{-1}\left(\vz_i - \vh_i(\vx) \right),
\end{equation}
where the inlier set $\cI_\textrm{inlier}  =  \{ \cI_\text{LOS}, \cI_\text{SB},\cI_\text{DB}\}$ is estimated as explained in Sections \ref{sec-snapshot} and \ref{sec-db-identify}.
It is important to note that since the data association between measurements and \acp{IP} has already been solved in steps 1-3 of the proposed algorithm, it is known which elements of $\vx$ correspond to $\cM_i$.

The problem defined above can be solved iteratively, for example, using the Gauss-Newton algorithm, which is based on approximating the measurement model using a first order Taylor series expansion around the linearization point $\hat{\vx}^{(j)}$ \cite{boyd2004convex}. The approximation is given by
\begin{equation}\label{eq-linearhx}
    \vh_i(\vx) \approx \vh_i(\hat{\vx}^{(j)}) +\vH_i(\hat{\vx}^{(j)})(\vx -\hat{\vx}^{(j)} ),
\end{equation}
in which the model $\vh_i(\hat{\vx}^{(j)}) \in \mathbb{R}^{d_{\vz} \times 1}$ and Jacobian $\vH_i(\hat{\vx}^{(j)}) = \partial \vh_i(\vx)/\partial \vx \in \mathbb{R}^{d_{\vz} \times d_\vx}$ are evaluated at the current linearization point $\hat{\vx}^{(j)}$.\footnote{Let us consider an example with linearization point $\hat{\vx}^{(j)} = [(\hat{\vs}^{(j)})^\top, (\hat{\vm}_1^{(j)})^\top, (\hat{\vm}_2^{(j)})^\top, (\hat{\vm}_3^{(j)})^\top, (\hat{\vm}_4^{(j)})^\top]^\top$ and double-bounce path with \acp{IP} $\hat{\cM}_i^{(j)} = \{ \hat{\vm}_2^{(j)}, \hat{\vm}_4^{(j)}\}$. The measurement model for $i \in \cI_\textrm{DB}$ is now defined as $\vh_i(\hat{\vx}^{(j)}) \triangleq \vh_i(\hat{\vs}^{(j)},\hat{\cM}_i^{(j)})$ and the Jacobian is
$
\vH_i(\hat{\vx}^{(j)}) = \begin{bmatrix}
    \vH_i(\hat{\vs}^{(j)}) & \mathbf{0}_{d_\vz \times d_\vm} & \vH_i(\hat{\vm}_2^{(j)}) &  \mathbf{0}_{d_\vz \times d_\vm} & \vH_i(\hat{\vm}_4^{(j)})
\end{bmatrix},
$
in which the partial derivatives $\vH_i(\hat{\vs}^{(j)})$, $\vH_i(\hat{\vm}_2^{(j)})$ and $\vH_i(\hat{\vm}_4^{(j)})$ are given in \eqref{eq:jacobian_db_ue}, \eqref{eq:jacobian_db_xl1} and \eqref{eq:jacobian_db_xl2}, respectively.} The update step of the Gauss-Newton algorithm can be derived by inserting \eqref{eq-linearhx} into \eqref{eq-objfunc}, setting the gradient of $L(\vx)$ to zero and solving for $\vx$. The gradient is given by
\begin{multline}
    \frac{\partial L(\vx)}{\partial \vx} \approx 2\sum_{i \in \cI_\textrm{inlier}}(\vH_i(\hat{\vx}^{(j)}))^\top \vR_i^{-1} \vH_i(\hat{\vx}^{(j)}) (\vx -\hat{\vx}^{(j)} ) \\ 
    -2\sum_{i \in \cI_\textrm{inlier}}(\vH_i(\hat{\vx}^{(j)}))^\top \vR_i^{-1} (\vz_i - \vh_i(\hat{\vx}^{(j)})),
\end{multline}
and solving for $\vx$ yields the updated state estimate
\begin{multline}
    \hat{\vx}^{(j+1)} = \hat{\vx}^{(j)} + \left( \sum_{i \in \cI_\textrm{inlier}}(\vH_i(\hat{\vx}^{(j)}))^\top \vR_i^{-1} \vH_i(\hat{\vx}^{(j)}) \right)^{-1} \\
    \times \left( \sum_{i \in \cI_\textrm{inlier}}(\vH_i(\hat{\vx}^{(j)}))^\top \vR_i^{-1} (\vz_i - \vh_i(\hat{\vx}^{(j)})) \right).
\end{multline}

The Gauss-Newton algorithm is initialized using the results obtained in steps $1-3$ of the proposed double-bounce snapshot \ac{SLAM} algorithm. The joint state of the \ac{UE} and \acp{IP} is initialized as $\hat{\vx}^{(0)} = [\hat{\vs}^\top, \hat{\vm}_{1}^\top,\cdots,\hat{\vm}_{N}^\top]^\top$ in which $\hat{\vs}$ is estimated as explained in Section \ref{sec-snapshot}, $N = \lvert \hat{\cM} \rvert$ is cardinality of the map, and $\hat{\vm}_{n} \in \hat{\cM}$ represent the \acp{IP} of the single and double-bounce paths that are estimated as explained in Sections \ref{sec-snapshot} and \ref{sec-closeform2IP}. The Gauss-Newton algorithm is run iteratively until the norm of the update step falls below a 
threshold ($\lVert \hat{\vx}^{(j+1)} - \hat{\vx}^{(j)} \rVert < T_\epsilon$) or a maximum number of iterations is reached ($j + 1 \geq J$).

\begin{figure*}[!t]
    \centering
    \includegraphics[trim={3cm 1.5cm 2cm 1.5cm},clip,width=\textwidth]{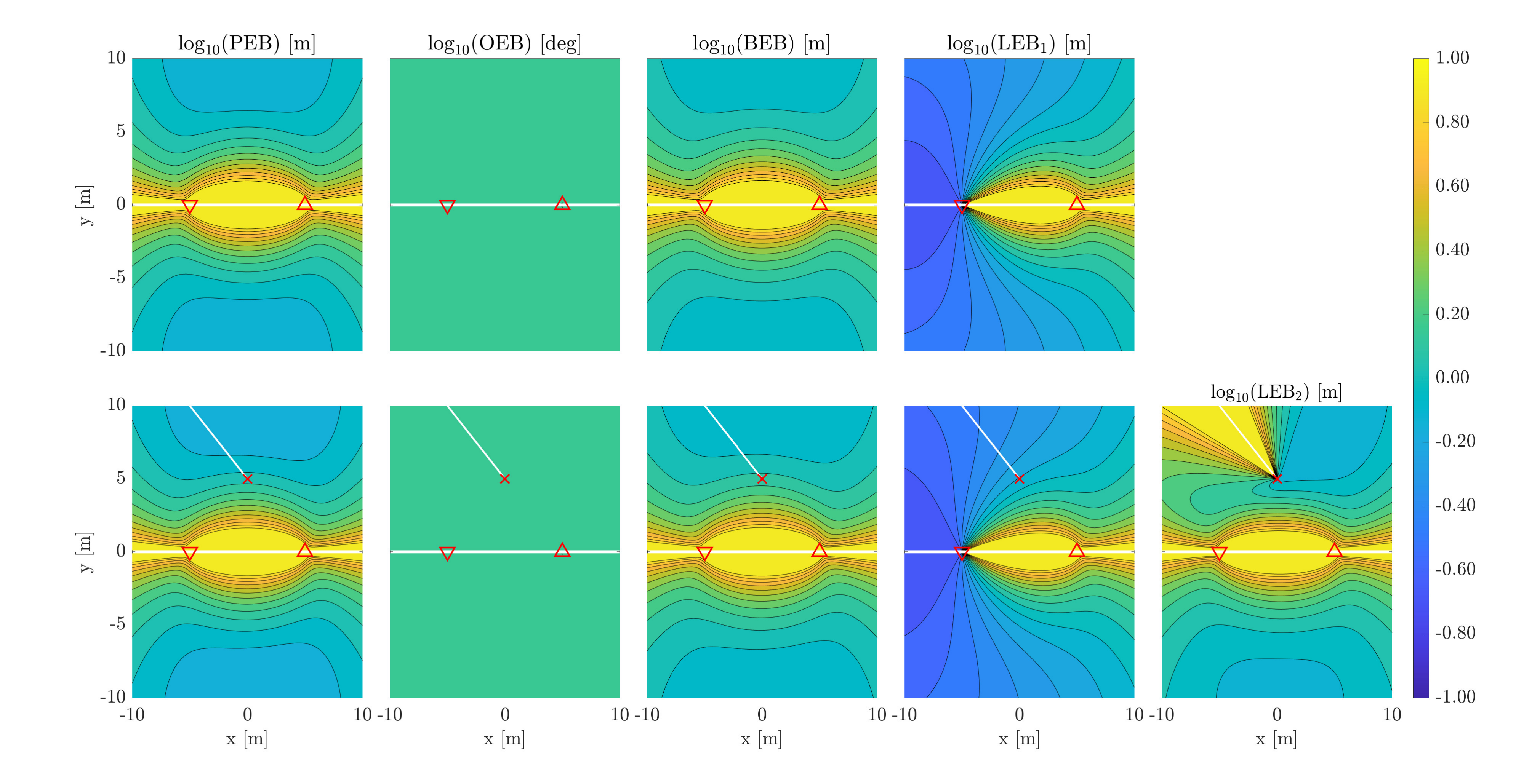}
    \vspace{-7mm}
    \caption{On top, the performance bounds with LoS and one single-bounce NLoS path. On bottom, the performance bounds with LoS, one single-bounce NLoS and one double-bounce NLoS path, which shares the first interaction point with the single-bounce NLoS path (case 2). In the figures, the IP of the single-bounce NLoS path is varied $x, \, y \in [-10\,\text{m},\, 10\,\text{m}]$, whereas the second IP of the double-bounce NLoS path is fixed to  $x = 0\,\text{m},\, y = 5\,\text{m}$. The BS location illustrated using  (\protect\redupsidedowntriangle), the UE location with (\protect\redtriangle), and the second IP of the double-bounce NLoS path using (\protect\redcross).}
    \label{fig:spatial_bounds_scenarios13}
    \vspace{-0mm}
\end{figure*}

\begin{figure*}[!t]
    \centering
    \includegraphics[trim={3cm 1.5cm 2cm 1.5cm},clip,width=\textwidth]{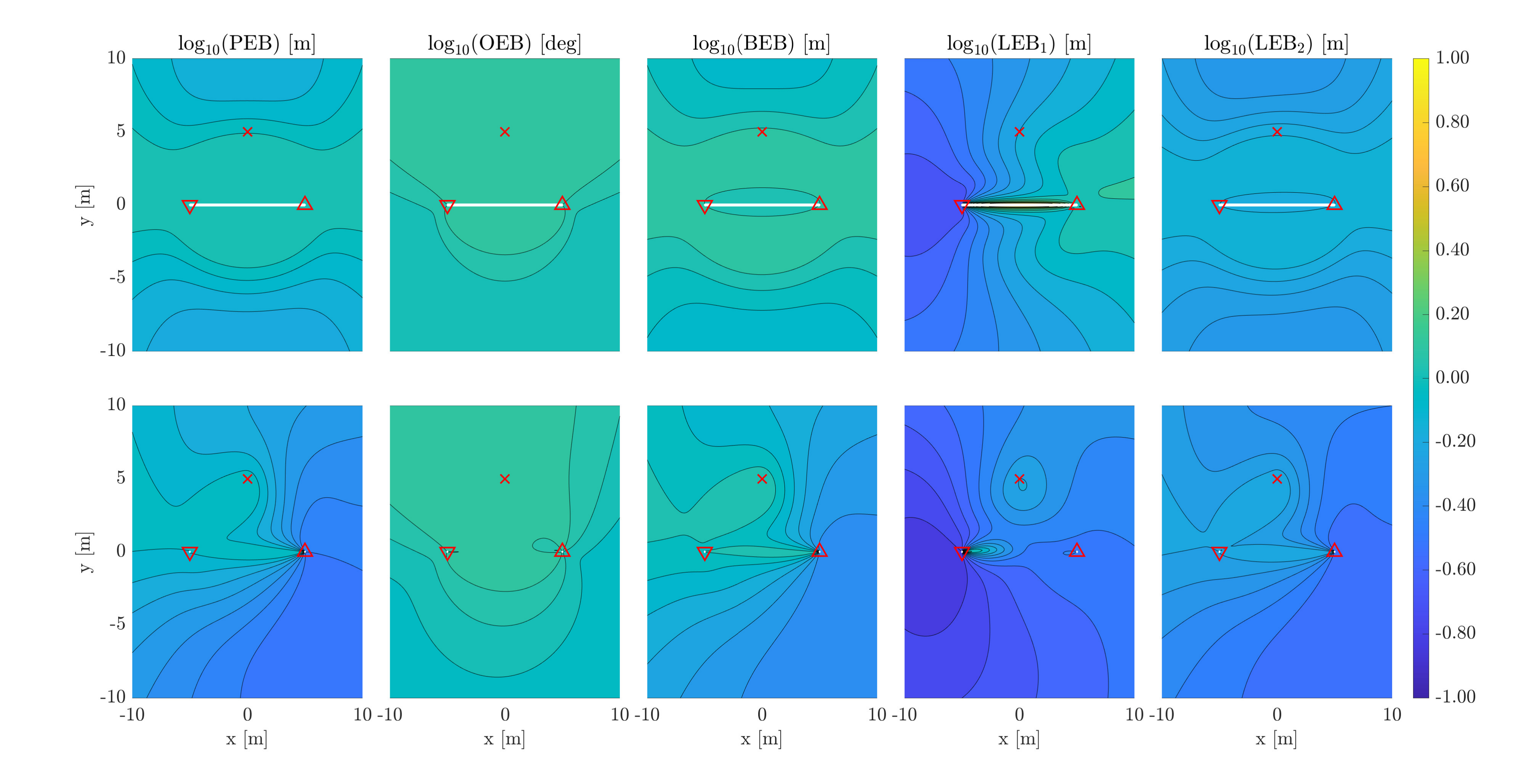}
     \vspace{-7mm}
    \caption{On top, the performance bounds with LoS and two single-bounce NLoS paths. On bottom, the performance bounds with LoS, two single-bounce NLoS paths and one double-bounce NLoS path, which shares both interaction points with the single-bounce NLoS paths (case 1). In the figures, the IP of one single-bounce NLoS path is varied $x_1, \, y_1 \in [-10\,\text{m},\, 10\,\text{m}]$, whereas the IP of the second single-bounce NLoS path is fixed to  $x_2 = 0\,\text{m},\, y_2 = 5\,\text{m}$. The double-bounce path propagates as follows: \ac{BS}~$\rightarrow [x_1, \, y_1]^\top \rightarrow [x_2, \, y_2]^\top \rightarrow$~\ac{UE}. The BS location illustrated using  (\protect\redupsidedowntriangle), the UE location with (\protect\redtriangle), and the IP of the fixed single-bounce NLoS path using (\protect\redcross).}
    \label{fig:spatial_bounds_scenarios24}
    \vspace{-2mm}
\end{figure*}

\vspace{-2mm}
\section{Results \label{sec-results-combined}}
\subsection{Numerical Evaluations and Insight \label{sec-numerical_results}}
\subsubsection{Simulation Setup}
The double-bounce \ac{SLAM} performance bounds are presented in the following. The analysis considers an example fixed \ac{BS} state, $x_\textrm{BS} = -5\,\text{m}, \, y_\textrm{BS} = 0\,\text{m}, \, \alpha_\textrm{BS} = 0\,\text{deg}$, fixed \ac{UE} state, $x_\textrm{UE} = 5\,\text{m}, \, y_\textrm{UE} = 0\,\text{m}, \, \alpha_\textrm{UE} = 90\,\text{deg}, \, b_\textrm{UE} = 0\,\text{ns}$, while the locations of the single-bounce and double-bounce \acp{IP} are bounded within the interval $x_n, \, y_n \in [-10\,\text{m},\, 10\,\text{m}]  \, \forall \, n$. It is assumed that the \ac{LOS} exists and covariance of the channel parameter estimates is set to $\vR_i= ([1 \text{ ns}, 1 \text{ deg}, 1 \text{ deg}]^2) \, \forall \, i$, which is the same as used in Section \ref{sec-experimental results}. In the following, we consider two different cases. In \textit{Case 1}, the double-bounce path shares both interaction points with the single-bounce  paths, whereas in \textit{Case 2}, the double-bounce path shares the first interaction point with the single-bounce path.

\subsubsection{Results and Discussion}
The single-bounce and double-bounce snapshot \ac{SLAM} performance bounds for \emph{Case 2} are illustrated in Fig.~\ref{fig:spatial_bounds_scenarios13}. In this figure, the top row presents the \ac{PEB}, \ac{OEB}, \ac{BEB}, and \ac{LEB} when one \ac{LOS} and one single-bounce path are available. The bottom row shows the corresponding bounds when a double-bounce path is present, sharing its first \ac{IP} with the single-bounce path. The \ac{LEB} for the second \ac{IP} of the double-bounce path is included in the last column as the double-bounce path introduces an additional \ac{IP} to the problem. It is evident that the problem geometry has a significant impact on the performance. In fact, the system is rank-deficient if the single-bounce \ac{IP} is on the line that traverses through the \ac{BS} and \ac{UE}, or the second double-bounce \ac{IP} is on the line segment connecting the first \ac{IP} and the \ac{UE}. These regions are illustrated with white lines in the figure and at these locations the problem is intractable since not all of the unknown parameters can be estimated. In such cases, however, it would be advantageous to either try to identify the paths that lead to the degenerate case so that they could be discarded,\footnote{The problem can still have full rank with a subset of paths. As an example, if the double-bounce path is discarded in Fig.~\ref{fig:spatial_bounds_scenarios13}, the problem can still be solved with the \ac{LOS} and single-bounce \ac{NLOS} for which the bounds are illustrated on the top row of the figure.} or use the Moore-Penrose inverse instead when the system does not have a unique solution. One of the aims of this paper is to highlight such rank-deficient problem geometries and developing robust estimator that can handle these special cases is left for future work. In general, the performance bounds for single- and double-bounce snapshot \ac{SLAM} are nearly identical, and on average the lower bound decreases only by $3.8\%$ for position, $0\%$ for orientation, $3.2\%$ for clock bias, and $5.0\%$ for the first landmark location when the double-bounce path is available. Thus, the main advantage of the double-bounce path in Case 2 is that estimation performance can be improved slightly, and that an additional landmark can be estimated which improves situational awareness of the surrounding environment, i.e., the accuracy and completeness of the overall map.

\begin{figure*}[!t]
    \centering
    \vspace{-5mm}
    \subfloat[PEB]{\includegraphics[width=0.4\textwidth]{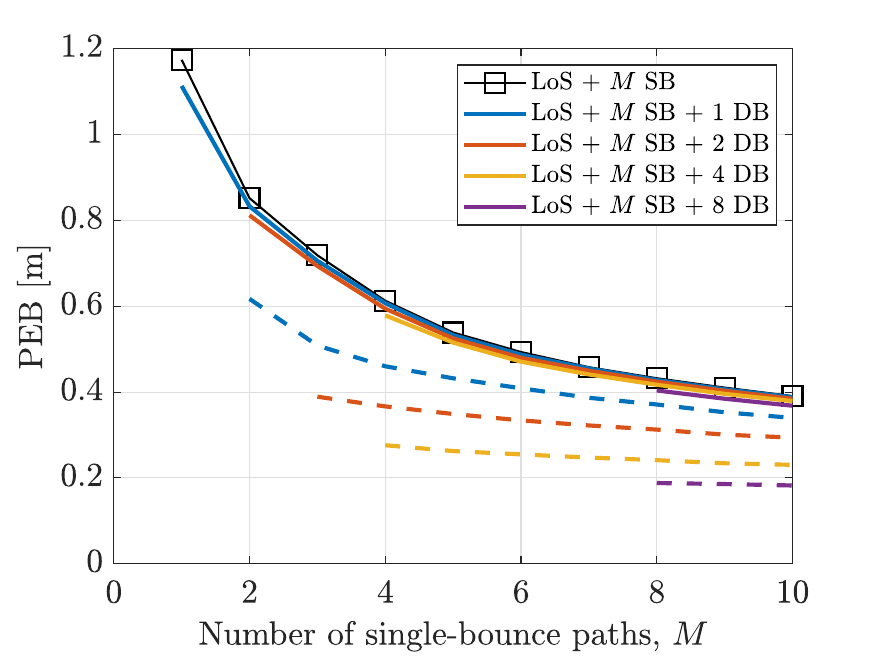}}
    \hfil
    \subfloat[LEB for the \acp{IP} of $M$ single-bounce paths]{\includegraphics[width=0.4\textwidth]{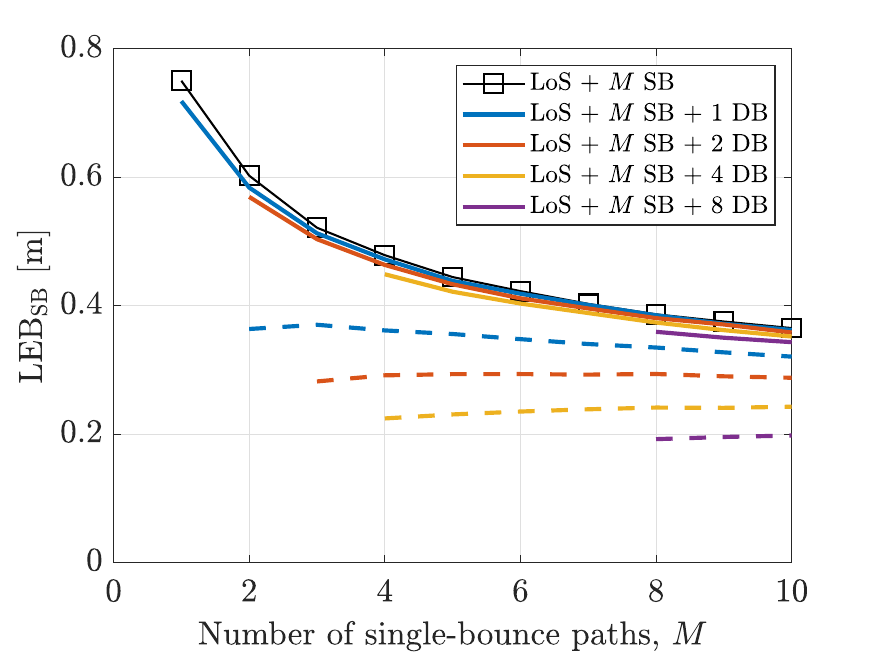}} \\
    
    \caption{PEB and LEB$_\text{SB}$ as a function of number of single-bounce paths (x-axis) and double-bounce paths (illustrated with different colors). In the figures, the scenario where every double-bounce path shares one IP with one of the single-bounce NLoS paths illustrated using solid lines (Case 2), whereas the dashed lines represent the scenario in which every double-bounce path shares both IPs with two different single-bounce NLoS paths (Case 1).} 
    \label{fig:bounds_vs_Npaths}
\end{figure*}

\textls[-3]{The performance bounds for \emph{Case 1} are illustrated in Fig.~\ref{fig:spatial_bounds_scenarios24}. In the top row, one \ac{LOS} path and two single-bounce paths are considered, and the \ac{PEB}, \ac{OEB}, \ac{BEB}, and \ac{LEB} for the two \acp{IP} are shown. In the bottom row, a double-bounce path that shares both \acp{IP} with the single-bounce paths is included, and the resulting bounds are presented. The inclusion of the double-bounce path breaks the symmetry properties of the \ac{PEB} and \ac{BEB}, as shown in the bottom row. This is due to the difference of the geometry: when the two \acp{IP} of the double-bounce path lie in different angular directions with respect to the \ac{BS} and \ac{UE}, the path can provide more spatial information compared to the case where both \acp{IP} lie along similar directions. As shown, compared to Fig.~\ref{fig:spatial_bounds_scenarios13}, the second single-bounce path provides additional information, thereby tightening the estimation bounds. Moreover, the double-bounce path provides three additional measurements (i.e., $\vz_i$) which introduce new system constraints without adding any new unknowns (i.e., $\vm_n$). As a result, the double-bounce path further improves the performance bounds, and on average the lower bound decreases by $36.8\%$ for position, $5.3\%$ for orientation, $35.3\%$ for clock bias, $44.1\%$ for the first landmark location, and $32.3\%$ for the second landmark location when the double-bounce path is available. Such improvements are really notable already. Interestingly, the double-bounce signal also improves identifiability of the system in between the \ac{BS} and \ac{UE}, since the $x$-coordinate of the first \ac{IP} becomes observable. To conclude, when the double-bounce path shares both of its \acp{IP} with the single-bounce \acp{IP}, the estimation performance enhances significantly and identifiability of the system also improves.}

Thus far, we have only considered a single double-bounce path and at most two single-bounce paths. In Fig.~\ref{fig:bounds_vs_Npaths}, \ac{PEB} and \ac{LEB}$_\text{SB}$ (LEB for the \acp{IP} of single-bounce paths) are illustrated as a function of the number of single- and double-bounce paths for two cases. As shown, the bounds are inversely proportional to the number of single-bounce paths, and the accuracy improves as the number of double-bounce paths increases. However, the two cases yield significantly different results. In Case 2, the improvement with respect to the single-bounce only scenario is marginal despite having up to eight double-bounce paths, whereas in Case 1, the performance improves significantly as the number of double-bounce paths increases. Interestingly in Case 1, the performance with three single-bounce paths and two double-bounce paths outperforms a system with ten single-bounce paths. The reason is that every single-bounce path provides  three additional known parameters and two additional unknowns, whereas the double-bounce paths do not increase the number of unknown parameters in Case 1. This implies that the double-bounce paths solely create additional constraints to the system which is beneficial in terms of estimation performance.


\vspace{-2mm}
\subsection{Experimental Results \label{sec-experimental results}}
\subsubsection{Experimental Setup}

\begin{figure*}[!t]
    \centering
    \subfloat[Single-bounce snapshot SLAM solution\label{fig:sub1}]{\includegraphics[trim={0.6cm 0cm 0.8cm 0.4cm},clip,width=0.32\textwidth]{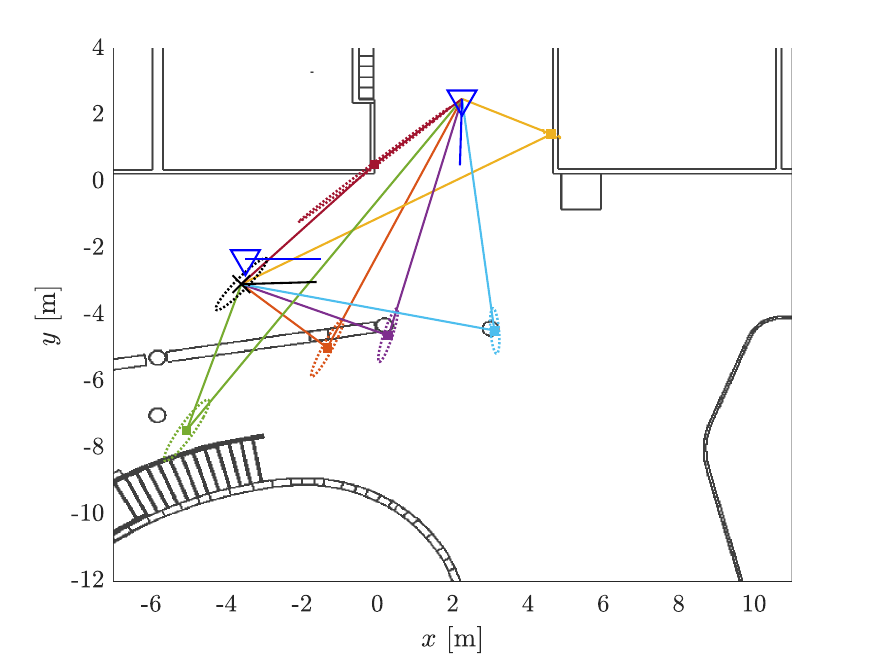}}
    \hfil
    \subfloat[Double-bounce classification\label{fig:sub2}]{\includegraphics[trim={0.6cm 0cm 0.8cm 0.4cm},clip,width=0.32\textwidth]{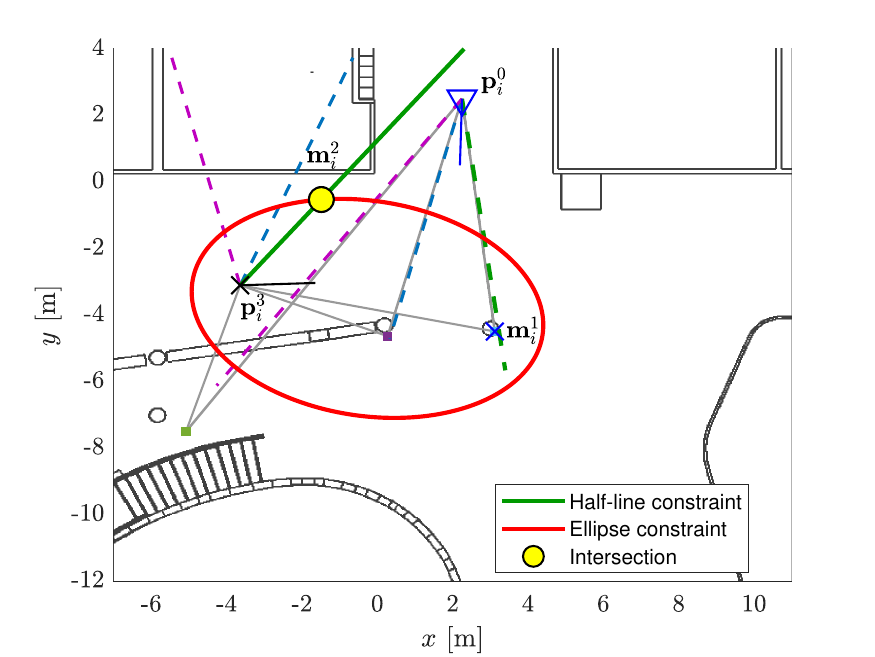}}
    \hfil
    \subfloat[Double-bounce snapshot SLAM solution\label{fig:sub3}]{\includegraphics[trim={0.6cm 0cm 0.8cm 0.4cm},clip,width=0.32\textwidth]{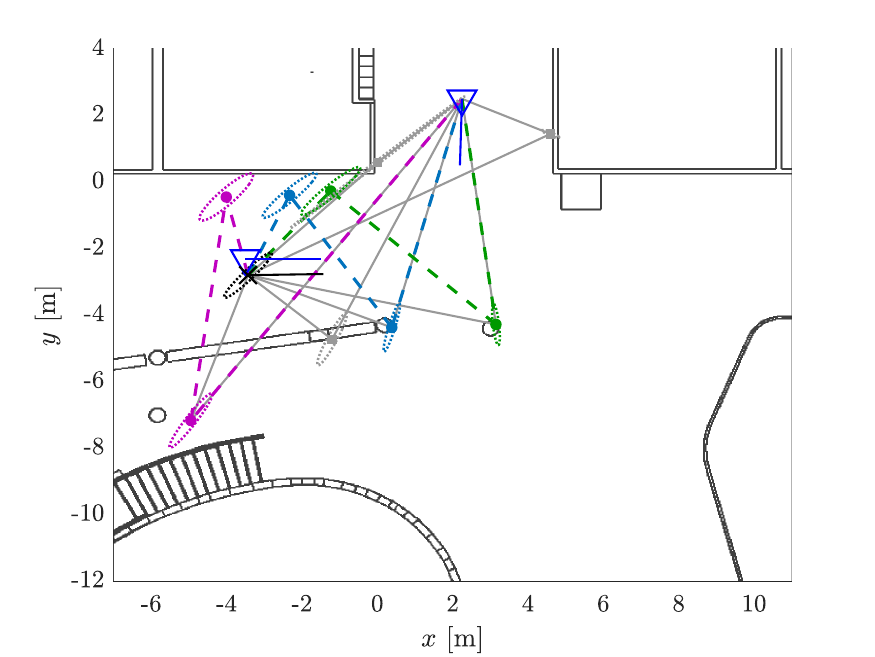}}
    \caption{Illustration at one example \ac{UE} position when the \ac{LOS} path does not exist. In the figures, ground truth \ac{BS} and \ac{UE} positions are shown using ($\textcolor{blue}{\bigtriangledown}$) and \ac{UE} position estimate using ($\times$). In the figures, the identified single-bounce paths illustrated using solid lines, double-bounce paths shown using dashed lines, whereas covariances of the \ac{UE} and \ac{IP} position estimates visualized with dotted ellipses. 
    In (a), the initial solution is shown which is computed only using the single-bounce paths. Figure (b) illustrates the steps of identifying and determining the double-bounce paths (outliers that are not classified as double-bounce signals are not visualized for clarity). In addition, estimating the second \ac{IP} is illustrated for one example double-bounce candidate. In (c), the final estimate is shown which is obtained using the proposed double-bounce snapshot SLAM algorithm.}
    \vspace{-2mm}
    \label{fig:example UE position}
\end{figure*}

We evaluate the performance of the proposed and benchmark \ac{SLAM} algorithms using an open access real-world 60 GHz measurement dataset, collected indoors on the campus of Tampere University \cite{mmwave_dataset}. In total, the dataset consist of 45 different \ac{UE} locations of which 32 were measured in \ac{LOS} conditions and 13 in \ac{NLOS} conditions. Beamformed measurements were obtained using 400 MHz transmission bandwidth utilizing 5G NR-specified downlink positioning reference signals. The experimental details and the hardware used are further explained in \cite{rastorgueva2024millimeter,mmwave_dataset}. For channel measurements, the \ac{AOD} and \ac{AOA} are estimated from the beam reference signal received power measurements using a constant false alarm rate-based detector and the \ac{TOA} is estimated using the singular value decomposition-based method introduced in \cite{rastorgueva2024millimeter}. The covariance of the channel parameters is empirically tuned and set to $\vR_i= ([1 \text{ ns}, 1 \text{ deg}, 1 \text{ deg}]^2) \forall i$. The angle threshold used to classify double-bounce paths in \eqref{eq:db_set_builder} and \eqref{eq:DB_condition} is set as $T_
{\vpsi} = 2 \text{ deg}$, the threshold to terminate the Gauss-Newton algorithm is $T_{\epsilon} = 0.1$, and the maximum number of allowed Gauss-Newton iterations is $J = 5$.


The proposed snapshot \ac{SLAM} algorithm is benchmarked with respect to two other snapshot \ac{SLAM} algorithms. The first benchmark is the robust snapshot \ac{SLAM} algorithm introduced in \cite{10818978}, which estimates the \ac{UE} state and landmarks using a \ac{LS} method. The second benchmark is the \ac{SLAM} algorithm introduced in \cite{zhang2025robust}, which instead applies \ac{MLE}. Both benchmarks solve the \ac{SLAM} problem by exploiting only the \ac{LOS} and single-bounce \ac{NLOS} paths; all the other measurements, including higher-order \ac{NLOS} paths are discarded. In this work, we refer to these benchmarks as SB-LS (single-bounce \ac{LS}) and SB-MLE (single-bounce \ac{MLE}), respectively. Performance is compared in terms of \ac{RMSE} of the estimated \ac{UE} position, orientation and clock bias. Ground-truth landmark positions are not available, which is a common problem using real-world experimental data.


\subsubsection{Results and Discussion}
\textls[-6]{Fig.~\ref{fig:example UE position} illustrates the sequential operation of the proposed estimator at one example \ac{UE} position under pure \ac{NLOS} condition. In the first stage, single-bounce paths are identified and an initial estimate of the \ac{UE} state together with single-bounce \acp{IP} is obtained as explained in Section~\ref{sec-snapshot}. The initial solution of the proposed algorithm, which is essentially the same as SB-LS, is visualized in Fig. \ref{fig:sub1}. In the example \ac{UE} position,  six single-bounce paths are identified and used to solve the problem. The position, heading and clock bias RMSE of the initial solution are [$0.82$ m, $2.00$ deg, $2.05$ ns], respectively. 
Building on the initial solution, Fig.~\ref{fig:sub2} depicts the classification and estimation of double-bounce paths for which the details are given in Section~\ref{sec-db-identify}. Three double-bounce candidates (dashed lines) are found whose \acp{AOD} align with that of existing single-bounce paths (solid lines). Consequently, for all three paths, the \ac{IP} of the aligned single-bounce path is treated as the first shared \ac{IP} of the double-bounce candidate. Once this shared \ac{IP} is determined, the second \ac{IP} of the double-bounce candidate can be estimated from the measurement constraints as explained in Section~\ref{sec-closeform2IP}. In the example of Fig.~\ref{fig:sub2}, the second \ac{IP} of the double-bounce path is estimated as the intersection between an ellipse whose foci are the \ac{UE} position and the first \ac{IP}, and a half-line that spans from the \ac{UE} toward the \ac{AOA} direction (green solid line). Lastly, both single-bounce and double-bounce paths are jointly exploited to refine the estimates, as shown in Fig.~\ref{fig:sub3}, and as explained in Section~\ref{sec:db_mle}. As illustrated, exploiting the three identified double-bounce paths allows mapping landmarks that are not visible with single-bounce paths alone. In addition, the double-bounce paths improve the accuracy since the position, heading and clock bias RMSE are reduced to [$0.50$ m, $0.96$ deg, $0.87$ ns], respectively. Thus, exploiting the double-bounce paths decreases the position, heading and clock error by $0.32$ m, $1.04$ deg, $1.18$ ns, respectively. Moreover, it is observable from the figures that the uncertainties of the \ac{UE} and existing \ac{IP} positions estimates are slightly reduced, since each double-bounce path adds information to the \ac{UE} and shared \acp{IP}, consistent with the \ac{FIM} analysis in Sec. \ref{sec:fim_double_bounce_path}.}

\begin{figure}[!t]
    \centering
    \includegraphics[trim={0.5cm 0.0cm 1.4cm 0.6cm},clip,width=0.8\columnwidth]{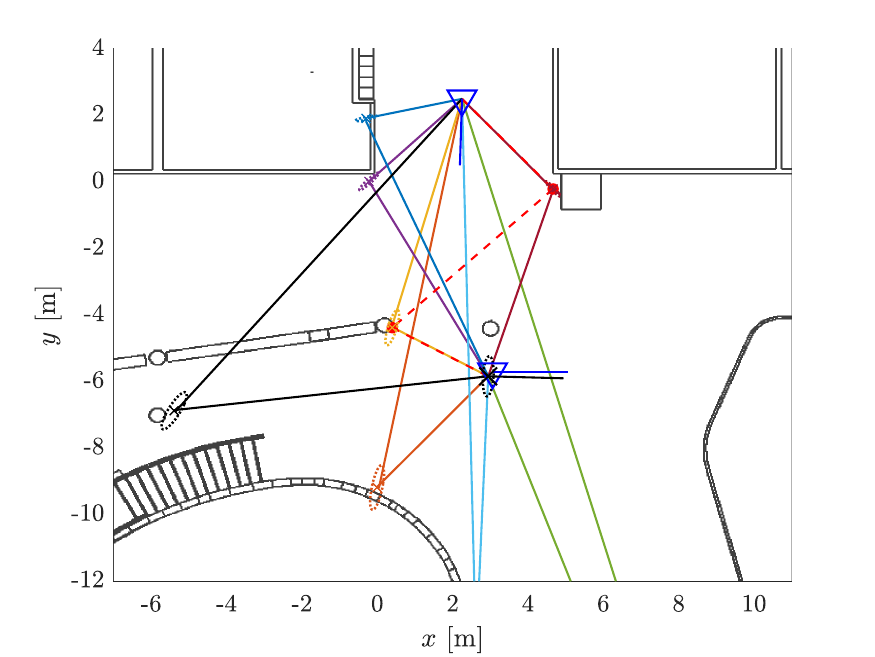}   
    \vspace{-2mm}
    \caption{Illustration of an example snapshot SLAM solution using the proposed algorithm when the double-bounce path shares both IPs with the single-bounce paths (shown using a red dashed line). }
    \label{fig:example UE results - both shared}
    \vspace{3mm}
\end{figure}

\begin{figure*}[!t]
    \centering
    \subfloat[Benchmark SB-LS\label{fig:sub4}]{\includegraphics[trim={0.3cm 0cm 0.8cm 0.6cm},clip,width=0.32\textwidth]{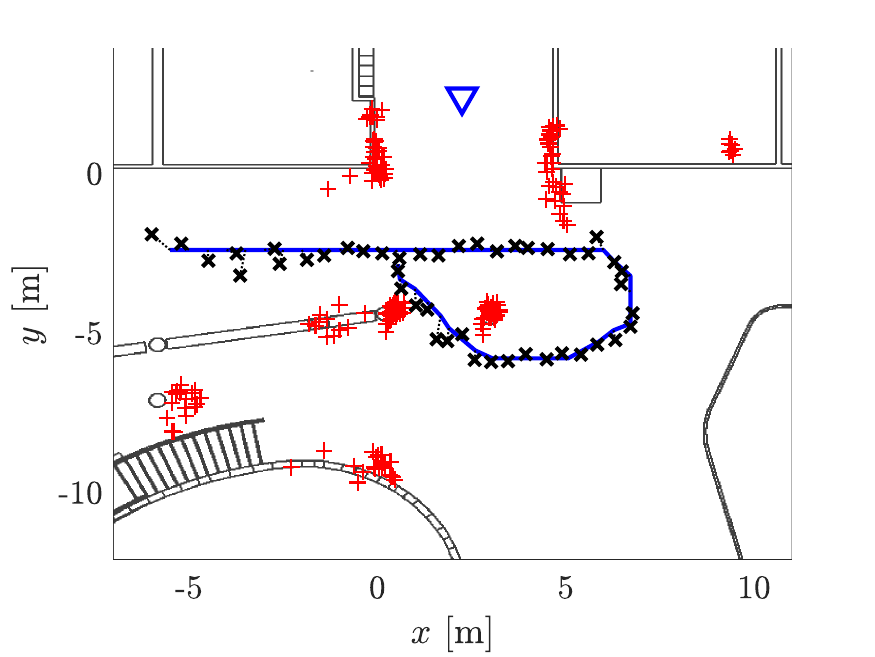}}
    \hfil
    \subfloat[Benchmark SB-MLE\label{fig:sub5}]{\includegraphics[trim={0.3cm 0cm 0.8cm 0.6cm},clip,width=0.32\textwidth]{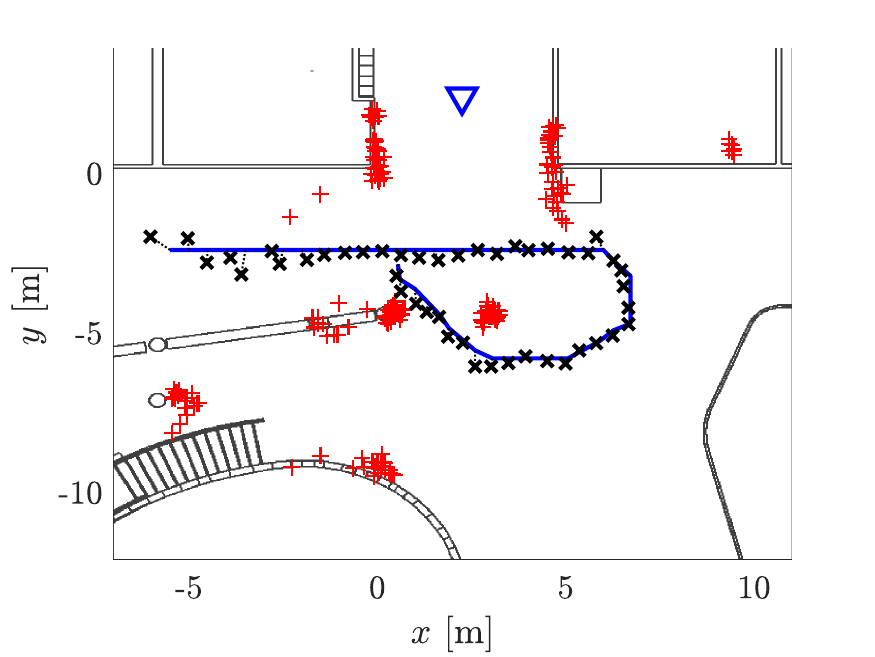}}
    \hfil
    \subfloat[Proposed DB-MLE\label{fig:sub6}]{\includegraphics[trim={0.3cm 0cm 0.8cm 0.6cm},clip,width=0.32\textwidth]{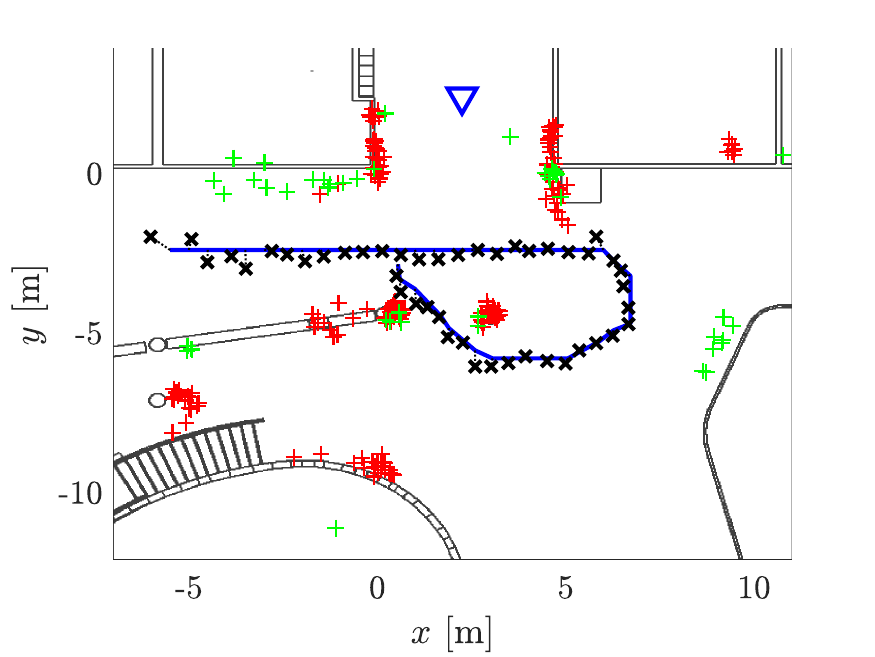}}
    \caption{SLAM performance using the benchmark algorithms and the proposed method. In the figures, BS position is shown using ($\textcolor{blue}{\bigtriangledown}$), the ground truth UE trajectory with ($\textcolor{blue}{\rule[0.5ex]{0.2cm}{1pt}}$), UE position estimates using ($\times$), single-bounce \ac{IP} position estimates with ($\textcolor{red}{+}$) and double-bounce \ac{IP} position estimates using ($\textcolor{green}{+}$).}
    \vspace{-1mm}
    \label{fig:visual performance}
\end{figure*}

\textls[-3]{Overall, a total of 424 resolvable propagation paths are collected in the 45 \ac{UE} positions. Among them, the benchmark methods and the initial solution of the proposed algorithm classify the paths as: 32 \ac{LOS} paths ($7.55 \%$), 233 single-bounce paths ($54.95\%$), and the remaining 159 ($37.5\%$) are categorized as multi-bounce paths and clutter. Using the proposed method with angular threshold $T_{\vpsi}=2\,\text{deg}$, 50 double-bounce \ac{NLOS} paths are detected among the remaining paths, accounting for $31.45\%$ of the subset. Notably, two of these double-bounce paths share both \acp{IP} with existing single-bounce paths. An example of such a case is shown in Fig. \ref{fig:example UE results - both shared}. In this case, adding the double-bounce path does not introduce any additional unknown parameters compared to SB-MLE. Instead, its measurements directly contribute to refining the estimates of the \ac{UE} state and the two shared \acp{IP}. The position, heading and clock bias RMSE of the benchmark and proposed methods are: [$0.26$ m, $2.23$ deg, $0.90$ ns] and [$0.18$ m, $1.73$ deg, $0.59$ ns], respectively.  Thus, the proposed method improves the \ac{UE} state estimates by $30.8\%$ for position, $22.4\%$ for orientation, and $34.4\%$ for clock bias, notably larger than the improvements observed when the double-bounce path shares only one \ac{IP} as we will discuss next.}

Fig.~\ref{fig:visual performance} visualizes the results over all \ac{UE} positions, and Table~\ref{t1-perf} summarizes the quantitative metrics. Across 45 \ac{UE} positions, the proposed method identifies a total of 50 double-bounce paths, spread over 32 \ac{UE} positions, while 2 double-bounce paths are detected that share both \acp{IP} with the single-bounce paths. Regarding environmental mapping, the proposed algorithm detects 48 additional landmarks, representing a $21\%$ increase relative to both single-bounce benchmarks. Particularly, double-bounce paths enable mapping of previous unobservable landmarks as illustrated in Fig.~\ref{fig:sub6}. In the figure,  green markers highlight additional walls and corners as shown in the upper-left and lower-right regions, that cannot be inferred from single-bounce paths alone. As summarized in Table \ref{t1-perf}, the overall \ac{RMSE} of the \ac{UE} state improves by $6.90\%$ for position, $2.98\%$ for orientation, and $10.20\%$ for clock bias compared with SB-MLE, and by $12.90\%$, $3.74\%$, and $20.72\%$, respectively, compared with SB-LS. Moreover, these gains come with no substantial increase in computational complexity since the main computation burden of the proposed and benchmark algorithms stem from the RANSAC-inspired search to identify single-bounce paths. Overall, the results are inline with the bound analysis conducted in Section~\ref{sec-numerical_results} which revealed that double-bounce signals improve the performance significantly more when the double-bounce path shares both of its \acp{IP} with the single-bounce \acp{IP}, compared to a situation where only one of the \acp{IP} is shared with a single-bounce \ac{IP}.

\begin{table}
\caption{RMSE performance comparison of the algorithms in terms of UE state estimation}
\begin{center}
\resizebox{0.45\textwidth}{!}{%
\begin{tabular}{|c|c|c|c|c|c|}
\hline
Methods   & Cond.             & Pos. [m] & Head. [deg] & Clk. [ns]& Time [ms]\\ \hline
Proposed  & \multirow{3}{*}{LoS ($32/45$)}  & $\mathbf{0.22}$         & $1.92$           & $\mathbf{0.80}$  &$1.79$    \\ 
SB-MLE   &        & $0.23$         & $\mathbf{1.91}$           & $0.81$  & $0.83$  \\
SB-LS &                       & $0.25$         & $1.93$           & $0.98$  & $0.65$     \\\hline
Proposed & \multirow{3}{*}{NLoS ($13/45$)}  & $\mathbf{0.35}$         & $\mathbf{2.03}$           & $\mathbf{1.05}$     &$689.17$   \\
SB-MLE &  & $0.41$         & $2.24$           & $1.31$  & $687.54$ \\ 
SB-LS &                       & $0.41$         & $2.27$           & $1.40$   &$682.80$\\\hline
Proposed & \multirow{3}{*}{All ($45/45$)}  & $\mathbf{0.27}$         & $\mathbf{1.95}$           & $\mathbf{0.88}$     &$200.37$    \\
SB-MLE  &  & $0.29$         & $2.01$           & $0.98$    &$199.21$ \\
SB-LS &                       & $0.31$         & $2.03$           & $1.11$  &$197.71$ \\ \hline
\end{tabular}%
}
\label{t1-perf}
\end{center}
\end{table}

As shown in Table~\ref{t1-perf}, in terms of the UE state estimation accuracy, the proposed method achieves average \ac{RMSE} improvements of [$0.03$ m, $0.01$ deg, $0.17$ ns] under \ac{LOS} conditions and [$0.06$ m, $0.24$ deg, $0.35$ ns] under \ac{NLOS} conditions, with respect to SB-LS. The gains under \ac{NLOS} conditions are on average higher. This trend is expected because, without a direct \ac{LOS} path, the \ac{UE} state is inferred solely from the \ac{NLOS} paths, which carry more correlated geometric information. Consequently, the benchmark leaves more room for improvement when additional paths are exploited under \ac{NLOS} conditions. Double-bounce paths contribute to these improvements by coupling the \ac{UE} state with two \acp{IP}, providing additional information. However, the magnitude of the improvement depends on factors such as the geometry at each snapshot and the number of double-bounce paths available. 



\vspace{-2mm}
\section{Conclusion \label{sec-conclusion}}
In this article, we addressed the role and potential of double-bounce \ac{NLOS} paths in snapshot \ac{SLAM} problem in terms of estimation performance and environmental mapping capabilities. First, we derived and analyzed the \ac{FIM} and the corresponding estimation \ac{CRB} of the snapshot \ac{SLAM} problem with double-bounce paths, and established generic identifiability conditions: the problem is solvable when the double-bounce paths share at least one \ac{IP} with existing single-bounce paths. Then, we proposed a new estimator framework and corresponding algorithms to identify double-bounce paths based on geometry consistency and further to solve the snapshot \ac{SLAM} problem. Numerical results on the bounds show that double-bounce paths sharing one \ac{IP} yield modest gains, whereas double-bounce paths sharing both \acp{IP} provide relatively larger improvements in estimation accuracy. Experimental results with the proposed estimator corroborate and support these fundamental trends: \ac{UE} estimation accuracy improves in line with the bound analysis. And most importantly, double-bounce paths also enable discovery and localization of new landmarks that are unobservable from single-bounce paths alone, expanding the maps without incurring significant computational overhead. Thus, the proposed methods provide an important step forward for extracting timely and accurate situational awareness of the physical environment, in the spirit of cellular ISAC, towards the 6G networks. Our future work will focus on extending the double-bounce methods to filtering and smoothing-based simultaneous localization and mapping, while also exploring the use of higher-order bounces and paths.
  
\vspace{-2mm}
\appendices

\section{Model Parameters and Jacobians}\label{sec:model_parameters_and_jacobian}

The parameters of the model in \eqref{eq:geometric_model} for \ac{LOS}, single-bounce \ac{NLOS} and double-bounce \ac{NLOS}, as well as the Jacobians $\vH_i(\vs) = \partial \vh_i(\vs, \cM_i)/ \partial \vs$ and $\vH_i(\vm_i) = \partial \vh_i(\vs, \cM_i)/ \partial \vm_i$ are defined in the following.
\subsubsection{LoS} The signal propagates directly from \ac{BS} to \ac{UE} ($\cM_i = \emptyset$). The model parameters are $d = \lVert \vp_\textrm{BS} - \vp_\textrm{UE} \rVert$, $[\delta_{\phi}^{x}, \, \delta_{\phi}^{y}]^\top = \vp_\textrm{BS} - \vp_\textrm{UE}$, $[\delta_{\theta}^{x}, \, \delta_{\theta}^{y}]^\top = \vp_\textrm{BS} - \vp_\textrm{UE}$ and the Jacobian is 
    \begin{equation}\label{eq:jacobian_los_ue}
        \vH_i(\vs) = \begin{bmatrix}
            -\delta_{\phi}^{x}/d & -\delta_{\phi}^{y}/d & 0 & 1 \\
            \delta_{\phi}^{y}/d^2 & -\delta_{\phi}^{x}/d^2 & 0 & 0 \\
            \delta_{\theta}^{y}/d^2 & -\delta_{\theta}^{x}/d^2 & -1 & 0
        \end{bmatrix}.
    \end{equation}

\subsubsection{Single-bounce NLoS} The signal experiences one interaction with the environment ($\cM_i = \{\vm_i^1\}$). The model parameters are $d = \lVert  \vp_\textrm{BS} - \vm_i^1 \rVert + \lVert \vm_i^1 - \vp_\textrm{UE} \rVert$, $[\delta_{\phi}^{x}, \, \delta_{\phi}^{y}]^\top =  \vp_\textrm{BS} - \vm_i^1$, $[\delta_{\theta}^{x}, \, \delta_{\theta}^{y}]^\top = \vm_i^1 - \vp_\textrm{UE}$ and the Jacobians are 
    \begin{align}
        \vH_i(\vs) &= \begin{bmatrix}
            -\delta_{\theta}^{x}/d_{\theta}  & -\delta_{\theta}^{y}/d_{\theta}  & 0 & 1 \\
            0 & 0 & 0 & 0 \\
            \delta_{\theta}^{y}/d_{\theta}^2 & -\delta_{\theta}^{x}/d_{\theta}^2 & -1 & 0
        \end{bmatrix}, \label{eq:jacobian_sb_ue} \\
        \vH_i(\vm_i^1) &= \begin{bmatrix}
            \delta_{\theta}^{x}/d_{\theta} - \delta_{\phi}^{x}/d_{\phi}   & \delta_{\theta}^{y}/d_{\theta} - \delta_{\phi}^{y}/d_{\phi} \\
            \delta_{\phi}^{y}/d_{\phi}^2 & -\delta_{\phi}^{x}/d_{\phi}^2 \\
            -\delta_{\theta}^{y}/d_{\theta}^2 & \delta_{\theta}^{x}/d_{\theta}^2 
        \end{bmatrix}, \label{eq:jacobian_sb_xl}
    \end{align}
    where $d_{\phi} = \lVert \vp_\textrm{BS} - \vm_i^1\rVert$ and $d_{\theta} = \lVert \vm_i^1 - \vp_\textrm{UE}\rVert$.

\subsubsection{Double-bounce NLoS} The signal experiences two interactions with the environment ($\cM_i = \{\vm_i^1, \,\vm_i^2\}$). The model parameters are 
    $d = \lVert \vp_\textrm{BS} - \vm_i^1 \rVert + \lVert \vm_i^2 - \vm_i^1  \rVert+ \lVert \vm_i^2 - \vp_\textrm{UE} \rVert$,
    $[\delta_{\phi}^{x}, \, \delta_{\phi}^{x}]^\top = \vp_\textrm{BS} - \vm_i^1$,
    $[\delta_{\theta}^{x}, \, \delta_{\theta}^{x}]^\top = \vm_i^2 - \vp_\textrm{UE}$,
    and the Jacobians are 
    \begin{align}
        \vH_i(\vs) &= \begin{bmatrix}
            -\delta_{\theta}^{x}/d_{\theta}  & -\delta_{\theta}^{y}/d_{\theta}  & 0 & 1 \\
            0 & 0 & 0 & 0 \\
            \delta_{\theta}^{y}/d_{\theta}^2 & -\delta_{\theta}^{x}/d_{\theta}^2 & -1 & 0
        \end{bmatrix}, \label{eq:jacobian_db_ue} \\
        \vH_i(\vm_i^1) &= \begin{bmatrix}
            \delta_{\varsigma}^{x}/d_{\varsigma} - \delta_{\phi}^{x}/d_{\phi}   & \delta_{\varsigma}^{y}/d_{\varsigma} - \delta_{\phi}^{y}/d_{\phi} \\
            \delta_{\phi}^{y}/d_{\phi}^2 & -\delta_{\phi}^{x}/d_{\phi}^2 \\
            0 & 0
        \end{bmatrix},  \label{eq:jacobian_db_xl1} \\
        \vH_i(\vm_i^2) &= \begin{bmatrix}
            \delta_{\theta}^{x}/d_{\theta} -\delta_{\varsigma}^{x}/d_{\varsigma}  & \delta_{\theta}^{y}/d_{\theta} -\delta_{\varsigma}^{y}/d_{\varsigma}  \\
            0 & 0\\
            -\delta_{\theta}^{y}/d_{\theta}^2 & \delta_{\theta}^{x}/d_{\theta}^2 
        \end{bmatrix}, \label{eq:jacobian_db_xl2}
    \end{align}
    where $[\delta_{\varsigma}^{x}, \, \delta_{\varsigma}^{y}]^\top = \vm_i^1 - \vm_i^2$, $d_{\varsigma} = \lVert \vm_i^1 - \vm_i^2\rVert$, $d_{\phi} = \lVert \vp_\textrm{BS} - \vm_i^1\rVert$ and $d_{\theta} = \lVert \vm_i^2 - \vp_\textrm{UE}\rVert$.

\bibliographystyle{IEEEtran} 
\bibliography{references}

\ifCLASSOPTIONcaptionsoff
  \newpage
\fi






\end{document}